\documentclass[aps,prb,amsmath,amssymb,amsfonts]{revtex4}
\usepackage[caption=false]{subfig}
\usepackage{graphicx}
\usepackage{color}
\usepackage{booktabs,array,multirow}

\begin{document}

\title{High-temperature tunable superfluidity of polaritons in Xene monolayers in an optical microcavity}
\author{Matthew N. Brunetti$^{1,2}$, Oleg L. Berman$^{1,2}$, and Roman Ya. Kezerashvili$^{1,2}$}
\affiliation{%
$^{1}$Physics Department,  New York City College of Technology\\
The City University of New York,
  300 Jay Street,   Brooklyn NY, 11201, USA \\
$^{2}$The Graduate School and University Center\\
The City University of New York,
New York, NY 10016, USA \\
}

\date{\today}

\begin{abstract}
  We study tunable polaritons in monolayers of silicene, germanene, and stanene (Xenes) via an external electric field in an open optical microcavity whose length can be adjusted.
  An external electric field applied perpendicular to the plane of the Xene monolayer simultaneously changes the band gap and the exciton binding energy, while the variable length of the open microcavity allows one to keep the exciton and cavity photon modes in resonance.
  First, the Schr\"{o}dinger equation for an electron and hole in an Xene monolayer is solved, yielding the eigenergies and eigenfunctions of the exciton as a function of the external electric field.
  The dependence of the polaritonic properties, such as the Rabi splitting and cavity photon damping, on the external electric field and on the cavity length, is analyzed.
  The Berezinskii-Kosterlitz-Thouless (BKT) transition temperature of polaritons is calculated as a function of the external electric field.
  We analyze and present the conditions for a room-temperature superfluid of lower polaritons by simultaneously maximizing the Rabi splitting and BKT transition temperature.
\end{abstract}

\maketitle

\section{\label{sec:intro}Introduction}

  Due to their dual nature as both matter and light, exciton-polaritons (hereafter: polaritons) exhibit a fascinating combination of light and matter properties, and they are therefore the ideal medium for studying a wide variety of quantum phenomena.
  For example, polaritons inherit their extremely small effective mass, $\approx 10^4~\text{m}_0$, from the effective mass of spatially confined photons~\cite{Snoke2017}, which significantly increases the superfluid critical temperature.
  It is also straightforward to experimentally detect and characterize polaritons since they couple directly to out-of-cavity photons with the same energy and in-plane wavevector~\cite{Snoke2017}.
  Meanwhile, the excitonic character of polaritons leads to polariton-polariton interactions which allows polaritons to thermalize, enabling the formation of quantum degenerate phases such as Bose-Einstein condensates (BEC)~\cite{Carusotto2013}.
  In addition, the photonic component enhances the phase-coherence of the spatial wavefunction of the polariton, which makes it robust against crystal defects which are usually fatal to the formation of a BEC of, for example, excitons~\cite{Deng2010}.
  In order to leverage these unique properties, polaritonic devices have been proposed for a wide variety of applications, from vertical-cavity surface-emitting lasers~\cite{Imamoglu1996,Malpuech2002,Christopoulos2007,Christmann2008}, to optical circuits~\cite{Miller2010,Shelykh2010,Gao2012a,Ballarini2013} and spin optical memory devices~\cite{Kavokin2007,Shelykh2008,Liew2008,Liew2010,Paraiso2010,Amo2010}.
  It can be seen, then, that polaritons are not just a physical curiosity but represent a very real path towards the development of next-generation optoelectronic devices.
  Comprehensive reviews of recent progress in polaritonic devices can be found in Refs.~\onlinecite{Liew2011,Sanvitto2016}.

  Amongst the most-sought-after phenomena in polaritonic devices is room-temperature superfluidity via the formation of a Bose-Einstein condensate of polaritons.
  In two-dimensional systems, the transition to the superfluid phase is characterized by the formation of bound vortex-anti vortex pairs, first described in the early 1970s by Berezinskii~\cite{Berezinskii1971,Berezinskii1972} and Kosterlitz and Thouless~\cite{Kosterlitz1972,Kosterlitz1973}.
  The Berezinskii-Kosterlitz-Thouless (BKT) phase transition model has been successfully used to describe the superfluid behavior of dilute, weakly interacting 2D Bose gases of excitons in semiconductor quantum wells (QWs)~\cite{De-Leon2001,Butov2004,Eisenstein2004,Berman2004,Berman2008d,Combescot2017,ProukakisSnoke2017}, gapped graphene~\cite{Berman2012b}, transition metal dichalcogenides (TMDCs)~\cite{Fogler2014,Wu2015,Berman2016,Berman2017a}, and phosphorene~\cite{Berman2017}, as well as polaritons in QWs~\cite{Snoke2002,Kavokinmalpeuchbook,Balili2007,Littlewood2007,Berman2008c,ProukakisSnoke2017} and gapped graphene~\cite{Berman2009,Berman2010c,Berman2012}.
  The first experimental evidence of a non-equilibrium polariton condensate was reported in Ref.~\onlinecite{Deng2002}, where the authors indirectly observed evidence of polariton condensation in GaAs/GaAlAs QWs at $T=4$ K; subsequent claims of polariton condensates followed~\cite{Deng2003,Richard2005}.
  Around the same time, theoretical works predicted that the BEC transition temperature was well above room-temperature for GaN~\cite{Malpuech2002}- and ZnO~\cite{Zamfirescu2002}-based microcavities.
  The first conclusive observation of polaritonic BEC was given in Ref.~\onlinecite{Kasprzak2006} in a CdTe multiple QW structure in an optical microcavity at $T=19$ K.
  Following this result, room-temperature condensation of polaritons was observed in the case of polariton lasing in bulk GaN in a microcavity~\cite{Christopoulos2007}.
  Superfluidity of polaritons was observed in Ref.~\onlinecite{Amo2009a}, albeit at $T=5$ K and with a Rabi splitting of only $5.1$ meV.
  Reviews of theoretical and experimental results on polariton condensation can be found in Refs.~\onlinecite{Keeling2007,Deng2010,Carusotto2013,Fraser2017,Snoke2017}.

  Since the advent of graphene, the atomically flat allotrope of carbon, in 2004~\cite{novoselov2004electric}, polaritonic research has begun to shift towards 2D materials.
  Atomically thin semiconductors such as TMDCs have several clear advantages over quasi-2D semiconductor QWs, namely their significantly enhanced exciton binding energies and extremely strong optical absorption by excitons when compared to semiconductor QWs~\cite{Mak2010,Eda2013,Xia2014,Wang2014,Brunetti2018} (see e.g. Refs.~\onlinecite{Wang2012,Geim2014,Mak2016,Wang2018} for reviews of the electronic and optical properties of TMDCs).
  Indeed, TMDCs have already been shown to exhibit the strong-coupling regime at room-temperature~\cite{Dufferwiel2015,Flatten2016}, and have been identified as candidates for room-temperature polaritonic devices~\cite{Lundt2017} and room-temperature superfluidity~\cite{Vasilevskiy2015}.
  TMDC bilayers and heterostructures consisting of TMDC monolayers sandwiching few-layer hexagonal boron nitride ($h$-BN) have also been identified as excellent candidates for high-temperature superfluidity of indirect (spatially separated) excitons~\cite{Fogler2014,Berman2016,Berman2017a}

  Another category of 2D semiconductors are the buckled 2D allotropes of silicon (Si), germanium (Ge), and tin (Sn), known as silicene, germanene, and stanene, and collectively referred to as Xenes~\cite{Molle2017} (for reviews of the properties of buckled 2D materials, see Refs.~\onlinecite{Kara2012,Houssa2015,Balendhran2015,Grazianetti2016,Chowdhury2016a}).
  Like TMDCs, Xenes exhibit very large exciton binding energies, but Xenes are unique amongst even the 2D materials because their buckled structure allows one to change the band gap using an external electric field aligned perpendicular to the Xene monolayer~\cite{Tabert2014}.
  By changing the band gap one also changes the effective mass of electrons and holes~\cite{Lalmi2010,Vogt2012,Drummond2012}, and therefore one can tune the binding energy and optical properties of excitons in Xenes using an electric field~\cite{Brunetti2018b}.

  In this paper, we study and examine the behavior and properties of polaritons in the freestanding Xenes, as well as in silicene encapsulated by $h$-BN, embedded in an open, variable-length microcavity.
  It is shown that Xenes are excellent candidates for extremely strong exciton-photon coupling and room-temperature superfluidity.
  In addition, we demonstrate that the exceptional tunability of Xenes via an external electric field, combined with the tunable nature of the open microcavity design, offers unprecedented control over the strength of the exciton-photon coupling, Rabi splitting, and BKT superfluid critical temperature.
  First, we determine the ground state properties of direct excitons formed in an Xene monolayer, namely the direct exciton binding energy, $E_b$, and excitonic Bohr radius, $a_B$, by solving the Schr\"{o}dinger equation for an interacting electron and hole constrained in the Xene monolayer plane, for which the interaction potential is the Rytova-Keldysh (RK) potential~\cite{Rytova1967,Keldysh1979}.
  We then obtain the properties of polaritons for an Xene monolayer embedded in a tunable-length open microcavity, in particular, the dependence of the Rabi splitting on the external electric field when the cavity length is changed in coincidence with the electric field so that the exciton and photon modes are kept in resonance.
  The theory of BEC and superfluidity in a 2D Bose gas of polaritons is then presented, and we examine the dependence of the BKT critical temperature on the external electric field for some fixed polariton concentration which is representative of a typical experimental setup.

  This paper is organized as follows.
  In Sec.~\ref{sec:theory}, we describe how an external electric field affects the band gap and binding energy of direct excitons in Xene monolayers.
  In Sec.~\ref{sec:cavity}, the properties of photons confined in an optical microcavity are presented.
  The Rabi splitting of polaritons in an open optical microcavity is calculated in Sec.~\ref{sec:polaritons}.
  In Sec.~\ref{sec:superfluidity}, the superfluid critical temperature of polaritons in an optical microcavity is calculated.
  The optimization problem of simulatenously maximizing the Rabi splitting and BKT critical temperature is analyzed in detail in Sec.~\ref{sec:opt}.
  We analyze our results and discuss their implications towards ongoing research in polaritons in 2D crystals and open microcavities in Sec.~\ref{sec:discussion}.
  Our conclusions follow in Sec.~\ref{sec:conc}.

  \section{\label{sec:theory}2D direct excitons in an Xene monolayer in an external electric field}

  Xene monolayers have a hexagonal lattice structure where the two triangular sublattices are vertically offset with respect to each other by a distance $d_0$, known as the buckling constant~\cite{Lalmi2010,Vogt2012}.
  If an electric field is applied perpendicular to the plane of the monolayer, a potential difference between the offset sublattices is created~\cite{Drummond2012}, which changes the band gap, and therefore the effective charge carrier masses, in the Xene monolayer.
  In the absence of an external electric field, the Xenes exhibit a prominent Dirac cone, though the Xenes have a small intrinsic gap ($\approx~1.9~\text{meV}$ in FS Si)~\cite{Matthes2013b}.
  The Hamiltonian of the band structure includes an additional term describing the dependence of the band gap on the external electric field~\cite{Kane2005,CastroNeto2009a,Abergel2010,Drummond2012}.
  The Fermi energy is set to the midway point between the valence band maximum (VBM) and the conduction band minimum (CBM), and the difference in energy between either the VBM or CBM and the Fermi energy is given by~\cite{Tabert2014}:

  \begin{equation}
	\Delta_{\xi \sigma} = \vert \xi \sigma \Delta_{\text{SO}} - e d_0 E_\perp \vert,
	\label{eq:deltaez}
  \end{equation}
  where $E_{\perp}$ is the perpendicular electric field, $\xi=\pm 1$ and $\sigma=\pm 1$ are the valley and spin indices, respectively, $\Delta_{\text{SO}}$ is half of the intrinsic band gap when $E_\perp = 0$, and $e$ is the electron charge.
  Eq.~\eqref{eq:deltaez} shows that the spin-up and spin-down valence and conduction bands are degenerate when $E_{\perp} = 0$.
  In other words, spin-orbit splitting only manifests itself at non-zero external electric fields.
  At non-zero electric fields, both the valence and conduction bands split, into ``upper'' bands with a large gap, (when $\xi\sigma = -1$), and ``lower'' bands with a small gap (when $\xi\sigma=1$).
  We refer specifically to the large band gap as $\Delta_{-1}$ and to the small band gap as $\Delta_{1}$.
  When the external field reaches a critical value $E_c = \Delta_{\text{SO}}/(e d_0)$, the ``lower'' bands form a Dirac cone at the $K/K'$ points.
  For $E_\perp \geq E_c$, both the ``upper'' and ``lower'' bands move away from the Fermi energy, and the difference in energy between the upper and lower conduction or valence bands is given by $2\Delta_{\text{SO}}$.
  This spin-orbit coupling gives rise to two types of excitons with different effective masses \textendash{} we refer to excitons formed from the ``large'' gap ($\xi\sigma = -1$) as $A$ excitons, and excitons formed from the ``small'' gap ($\xi\sigma=1$) as $B$ excitons.

  In the vicinity of the K/K' points, the conduction and valence bands are parabolic.
  Writing the dispersion relation as $E(\mathbf{k}) = \sqrt{\Delta_{\xi \sigma}^{2} + \hbar^2 v_F^2 \mathbf{k}^2}$~\cite{Tabert2014}, where $v_F$ is the Fermi velocity, and performing a Taylor expansion for small $\mathbf{k}$, we can identify the effective mass of charge carriers as $m = \Delta_{\xi\sigma}/v_F^2$.
  Due to the symmetry of the conduction and valence bands, the effective masses of electrons and holes are the same, $m_h = m_e = m$, and can be written as:

  \begin{equation}
	m = \frac{\lvert \xi\sigma\Delta_{\text{SO}} - e d_0 E_\perp \rvert}{v_F^2}.
	\label{eq:effmassEz}
  \end{equation}

  Therefore, both the band gap~\eqref{eq:deltaez} and effective carrier mass~\eqref{eq:effmassEz} depend the external electric field, demonstrating that it is imperative that one also obtains accurate values for the quantities $\Delta_{\text{SO}}$, $d_0$, and $v_F$.

  The eigenenergies and eigenfunctions of the exciton can be obtained by solving the 2D Schr\"{o}dinger equation.
  We treat the electron-hole interaction using the Rytova-Keldysh (RK) potential~\cite{Rytova1967,Keldysh1979}, which has been widely used to describe the screened electron-hole interaction in different 2D materials~\cite{Berkelbach2013,VanDerDonck2017,Brunetti2018b}.
  The RK potential is given by:

  \begin{equation}
	V_{\text{RK}}(\mathbf{r}) = \frac{\pi k e^2}{2 \kappa \rho_0} \left[ H_0 \left( \frac{\mathbf{r}}{\rho_0} \right) - Y_0 \left( \frac{\mathbf{r}}{\rho_0} \right) \right],
	\label{eq:rkpot}
  \end{equation}
  where $\mathbf{r} = \mathbf{r}_e - \mathbf{r}_h$ is the electron-hole separation, $\rho_0 = \left( l \epsilon \right)/\left( 2 \kappa \right)$ is the screening length, $l$ is the thickness of the Xene monolayer, $\epsilon$ is the static dielectric constant of the Xene monolayer, $\kappa = (\epsilon_1 + \epsilon_2)/2$, with $\epsilon_1$ and $\epsilon_2$ denoting the dielectric constants of the materials above and below the Xene monolayer, and $H_0$ and $Y_0$ are the Struve and Bessel functions of the second kind, respectively.

  After separation of the center-of-mass and relative motion, the Schr\"{o}dinger equation for the exciton reads:

  \begin{equation}
	\left[ \frac{- \hbar^2}{2 \mu} \nabla^2 + V_{\text{RK}} \left( \mathbf{r} \right) \right] \psi\left( \mathbf{r} \right) = E \psi \left( \mathbf{r} \right),
	\label{eq:relschro}
  \end{equation}
  where $\mu = (m_e m_h)/(m_e + m_h) = m/2$ is the reduced mass of the exciton.

  A detailed study of the field-dependent excitonic properties in Xene monolayers and Xene/$h$-BN heterostructures based on the solution of Eq.~\eqref{eq:relschro} was performed in Ref.~\onlinecite{Brunetti2018b}.
  Notably, these calculations demonstrated that the freestanding Xenes exhibit a phase transition from the excitonic insulator phase to the semiconductor phase as the electric field is increased beyond some critical value $E_{\perp,c}$, which was addressed in more detail in Ref.~\onlinecite{Brunetti2018c}.
  Therefore, we will only consider the formation of polaritons in the freestanding Xenes for $E_{\perp} > E_{\perp,c}$.

\section{\label{sec:cavity}Microcavity Properties and Parameters}

  We consider a microcavity containing an Xene monolayer placed on top of a DBR mirror.
  The opposite end of the cavity comprises a movable stage which enables adjustment of the cavity length, $L_c$.
  Throughout the following calculations we consider that the cavity length is changed in coincidence with the electric field $E_\perp$ such that the excitonic and photonic modes remain in resonance.
  The cavity length determines the allowed resonant photon modes, $E_{ph}$, and the energy of these modes is related to $L_c$ as:

  \begin{equation}
	E_{ph} = \frac{\hbar \pi c}{L_c \sqrt{\epsilon_{cav}}}.
	\label{eq:Lc}
  \end{equation}

  The mirror placed on the movable stage can be either another DBR~\cite{Dufferwiel2014,Dufferwiel2015,Wang2016}, or a metallic mirror~\cite{Flatten2016}.
  A crucial difference between the choice of a DBR or metallic mirror is that a DBR introduces additional photonic path length since reflection from a DBR involves the photon penetrating some distance into the DBR.
  For each DBR, the additional photonic path length is given by~\cite{Savona1995,Ciuti1998,Savona1999,Dufferwiel2014}:

  \begin{equation}
	  L_{\text{DBR}} = \frac{\lambda_c}{2 \sqrt{\epsilon_{cav}}} \frac{n_1 n_2}{(n_2 - n_1)} = \frac{h c}{2 E_{ph} \sqrt{\varepsilon_{cav}}} \frac{n_1 n_2}{(n_2 - n_1)},
	  \label{eq:Ldbr}
  \end{equation}
  where $n_1$ and $n_2$ are the refractive indices of the dielectrics composing the DBR, and $\lambda_c$ is the central wavelength of the DBR, at which the mirror is maximally reflective.
  Therefore, the effective photonic path length of the microcavity can be written $L_{eff} = L_c + L_{\text{DBR}}$ for the case of a microcavity with one DBR and one silver mirror, and $L_{eff} = L_c + 2 L_{\text{DBR}}$ if the microcavity consists of 2 identical DBR mirrors.
  Assuming a typical DBR consisting of alternating layers of SiO$_2$ $(n_1 = 1.45)$ and TiO$_2$ $(n_2 = 2.05)$~\cite{Flatten2016}, where $\lambda_c$ corresponds to the $E_{ph}$ given by a particular $L_c$, we find that $L_{\text{DBR}}$ is greater than $L_c$ by nearly a factor of 4, so the choice between a 2 DBR and 1 DBR microcavity configuration significantly changes the value of $L_{eff}$.

  In an optical microcavity, the rate of photon leakage from the microcavity can be considered a type of damping in the system, and the contribution to the photonic damping due to the leakage from a single mirror, $\gamma_{ph}^{'}$, is given by~\cite{Flatten2016}:

  \begin{equation}
	\gamma_{ph}^{'} = \frac{1 - \sqrt{R}}{\sqrt{R}} \frac{c}{\sqrt{\epsilon_{cav}} (2 L_{eff})},
	\label{eq:gamma1mirror}
  \end{equation}
  where $R$ is the reflectivity of the mirror, and the second fraction represents the time it takes for the photon to travel back and forth across the cavity once.
  Therefore, the decay rate of photons from the cavity is given by the sum of the photonic decay rate from each mirror:

  \begin{align}
	\gamma_{ph}^{\text{(2 DBR)}}	& = \frac{1 - \sqrt{R_1}}{\sqrt{R_1}} \frac{c}{\sqrt{\epsilon_{cav}} L_{eff}},\\
	\gamma_{ph}^{\text{(1 DBR)}}	& = \left[ \frac{1}{2} \left( \frac{1 - \sqrt{R_1}}{\sqrt{R_1}} + \frac{1 - \sqrt{R_2}}{\sqrt{R_2}} \right) \right] \frac{c}{\sqrt{\varepsilon_{cav}} L_{eff}} \nonumber \\
									& = \frac{1 - \sqrt{R_{eff}}}{\sqrt{R_{eff}}} \frac{c}{\sqrt{\varepsilon_{cav}} L_{eff}},%
	\label{eq:gamcav}
  \end{align}
  for each microcavity configuration considered in this work.
  In Eq.~\eqref{eq:gamcav}, $R_1$ is the reflectivity of the DBR, $R_2$ is the reflectivity of the metallic mirror, and $R_{eff} = (4 R_1 R_2)/(\sqrt{R_1} + \sqrt{R_2})^2$ is the effective reflectivity of a microcavity with two non-identical mirrors.

  In this work we consider two open ($\epsilon_{cav} = 1$) microcavity configurations, one consisting of two DBRs and one with one DBR and one silver mirror.
  The microcavity is characterized by the following parameters: a DBR consisting of SiO$_2$/TiO$_2$~\cite{Flatten2016} with refractive indices $n_1 = 1.45$ and $n_2 = 2.05$, respectively, and reflectivity $R_1 = 0.985$~\cite{Weisbuch1992,Savona1995}; the silver mirror has a reflectivity $R_2 = 0.95$~\cite{Flatten2016}.

\begin{figure}[h]
	\centering
	\includegraphics[width=0.60\columnwidth]{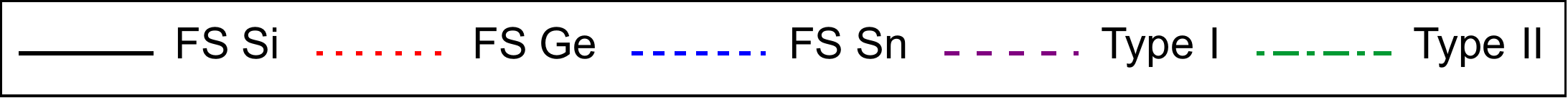}
	\includegraphics[width=0.49\columnwidth]{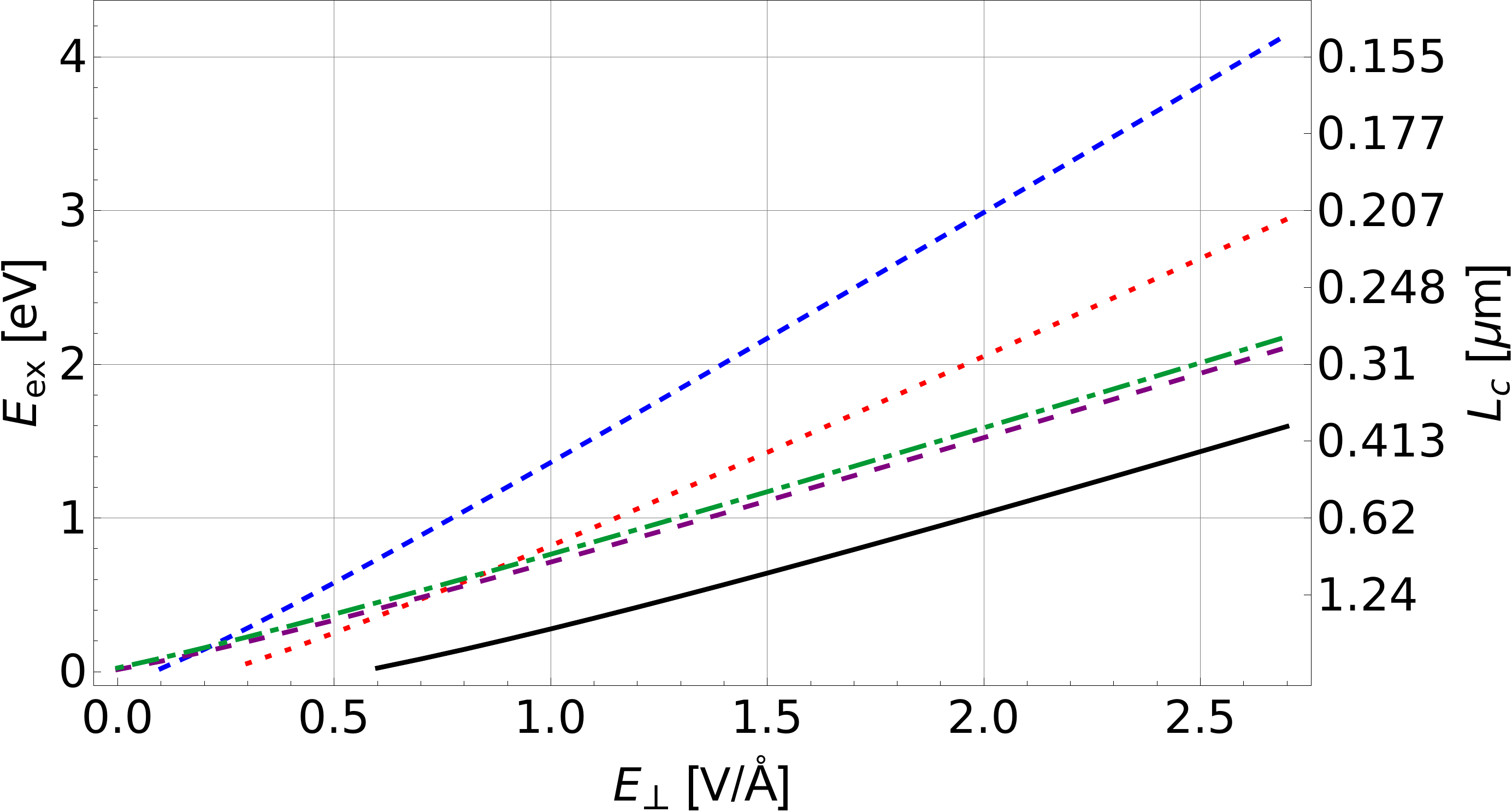}
	\caption{%
	  Spectrum of $A$ exciton transition energy, $E_{ex,A} = (2\Delta_{-1} - E_{b,A})$, at $\mathbf{k}=0$ as a function of the electric field $E_{\perp}$.
	  The right hand side of the frame shows the correspondence between the exciton energy $E_{ex}$ and the cavity length $L_c$ necessary to keep $E_{ph} = E_{ex}$.
	}\label{fig:exresonance}
  \end{figure}

  We calculated the dependence of the $A$ exciton transition energy, $E_{ex,A}$, on the external electric field $E_{\perp}$, for each of the 5 materials under consideration, and present our results in Fig.~\ref{fig:exresonance}.
  Also shown along the right side of the frame is the cavity length $L_c$ which yields a photon energy in resonance with the excitonic transition energy, $E_{ph} = E_{ex}$.

\section{\label{sec:polaritons}Polaritons in an Optical Microcavity}

  The Hamiltonian of the exciton-photon interaction is given by~\cite{Kavokinbook}:

  \begin{equation}
	\hat{H}_0 = \sum_{\mathbf{k}} E_{ex} (\mathbf{k}) \hat{b}_{\mathbf{k}}^{\dag} \hat{b}_{\mathbf{k}} +
	\sum_{\mathbf{k}} E_{ph}(\mathbf{k}) \hat{a}_{\mathbf{k}}^{\dag} \hat{a}_{\mathbf{k}} +
	\hbar V \sum_{\mathbf{k}} \left( \hat{a}_{\mathbf{k}}^{\dag} \hat{b}_{\mathbf{k}} + \hat{b}_{\mathbf{k}}^{\dag} \hat{a}_{\mathbf{k}} \right) +
	\frac{1}{2A} \sum_{\mathbf{k},\mathbf{k}',\mathbf{q}} U_{\mathbf{q}} b^{\dag}_{\mathbf{k}'+\mathbf{q}} b^{\dag}_{\mathbf{k}-\mathbf{q}}b_{\mathbf{k}}b_{\mathbf{k}'},
	\label{eq:cavityham}
  \end{equation}
  where $\hat{a}_{\mathbf{k}}$ $(\hat{a}_{\mathbf{k}}^{\dag})$ and $\hat{b}_{\mathbf{k}}$ $(\hat{b}_{\mathbf{k}}^{\dag})$ are the photonic and excitonic Bose annihilation (creation) operators, respectively.

  The first term in Eq.~\eqref{eq:cavityham} is the Hamiltonian of non-interacting excitons, where

  \begin{equation}
	E_{ex} (\mathbf{k}) = E_{ex} + \frac{\hbar^2 \mathbf{k}^2}{2 M_{ex}}
	\label{eq:excdisp}
  \end{equation}
  is the dispersion relation of a single exciton in the Xene monolayer with in-plane momentum $\mathbf{k}$, and $M_{ex} = 2 m$ is the total mass of the exciton.

  The second term in Eq.~\eqref{eq:cavityham} is the Hamiltonian of non-interacting photons confined in a semiconductor microcavity~\cite{Pau1995}, where

  \begin{equation}
	E_{ph} (\mathbf{k}) = \frac{\hbar c}{\sqrt{\epsilon_{cav}}} \sqrt{\frac{\pi^2}{L_c^2} + \mathbf{k}^2}
	\label{eq:photdisp}
  \end{equation}
  is the dispersion relation of the photon~\cite{Berman2012}.
  Assuming $\mathbf{k}$ is small, Eq.~\eqref{eq:photdisp} can be expanded to obtain~\cite{Deng2010}

  \begin{equation}
	E_{ph} \left( \mathbf{k} \right) \approx E_{ph} + \frac{\hbar^2 \mathbf{k}^2}{2 m_{ph}},
	\label{eq:photdispapprox}
  \end{equation}
  where $m_{ph} = \left( E_{ph} \epsilon_{cav} \right)/\left( c^2 \right)$ is the effective mass of the photon due to confinement within the cavity.

  The third term in Eq.~\eqref{eq:cavityham} is the Hamiltonian of harmonic exciton-photon coupling~\cite{Ciuti2003}, where $V$ is the exciton-photon coupling constant.
  The functional form of $V$ depends on the system in question, but following Refs.~\onlinecite{Savona1995,Dufferwiel2015,Vasilevskiy2015,Deveaudbook}, the exciton-photon coupling constant in this system can be written as:

  \begin{equation}
	V = \left[ N_X \frac{1 + \sqrt{R}}{\sqrt{R}} \frac{4 \pi k e^2 v_F^2}{E_{ex} L_{eff} \sqrt{\varepsilon_{cav} \kappa}} \lvert \psi(0) \rvert^2 \right]^{1/2},
	\label{eq:SavVplug}
  \end{equation}
  where $N_X$ is the number of Xene monolayers in the microcavity, $R$ is the reflectivity of the mirrors, $L_{eff}$ is the effective cavity length, $\psi(0)$ is the value of the exciton relative motion wavefunction evaluated at $r=0$, and the expressions for $R$ and $L_{eff}$ are determined by the choice of either the 1 DBR or 2 DBR microcavity setup.
  A brief summary of how Eq.~\eqref{eq:SavVplug} is obtained is given in Appendix~\ref{app:v}.

  The fourth term in Eq.~\eqref{eq:cavityham} describes the repulsive exciton-exciton interaction potential.
  As a first step, we neglect this term while considering the formation of polaritons in the microcavity.

  The eigenenergies of Eq.~\eqref{eq:cavityham} can be obtained by diagonalizing the Hamiltonian using a well-established procedure~\cite{Hopfield1958} (see Appendix~\ref{app:hopcoeff}).
  To properly account for excitonic and photonic damping, which originate from the finite linewidth of the excitonic transition and the leakage rate of photons from the mirrors, we write $E_{ex}$ and $E_{ph}$ as explicity complex, that is, $E_{ex} \to E_{ex} - i \hbar \gamma_{ex}$ and $E_{ph} \to E_{ph} - i \hbar \gamma_{ph}$.
  Then the complex upper/lower polariton eigenergies are given by~\cite{Deng2010,Kavokin2007}

  \begin{equation}
	E_{\text{UP}/\text{LP}}(\mathbf{k}) = \frac{E_{ph}(\mathbf{k}) + E_{ex}(\mathbf{k}) - i \hbar \left( \gamma_{ex} + \gamma_{ph} \right)}{2} \pm \sqrt{\hbar^2 V^2 + \frac{1}{4} \left[ \Delta E \left( \mathbf{k} \right) + i \hbar \left( \gamma_{ex} - \gamma_{ph} \right) \right]^2},
	\label{eq:ULpoldamping}
  \end{equation}
  where $\Delta E (\mathbf{k}) = E_{ph} \left( \mathbf{k} \right) - E_{ex} \left( \mathbf{k} \right)$ is the so-called detuning between the bare exciton and photon modes.
  The real and imaginary parts of the complex eigenenergies of Eq.~\eqref{eq:ULpoldamping} correspond to the eigenenergies, $E_{\text{UP}}$ and $E_{\text{LP}}$, and decay rates, $\gamma_{\text{UP}}$ and $\gamma_{\text{LP}}$, of upper and lower polaritons, respectively.
  The polariton decay rates can also be calculated directly as~\cite{Deng2010},
  \begin{align}
	\gamma_{\text{LP}}(\mathbf{k}) & = \lvert X_{\mathbf{k}} \rvert^2 \gamma_{ex} + \lvert C_{\mathbf{k}} \rvert^2 \gamma_{ph} \nonumber \\
	\gamma_{\text{UP}}(\mathbf{k}) & = \lvert C_{\mathbf{k}} \rvert^2 \gamma_{ex} + \lvert X_{\mathbf{k}} \rvert^2 \gamma_{ph},
	\label{eq:pollifetime}
  \end{align}
  where $\lvert X_{\mathbf{k}} \rvert $ and $\lvert C_{\mathbf{k}} \rvert $ are the Hopfield coefficients of Eq.~\eqref{eq:hopfielddefs}, but it turns out that calculating $\gamma_{\text{UP/LP}}\left( \mathbf{k} \right)$ using Eq.~\eqref{eq:pollifetime} yields exactly the same result as taking the imaginary part of Eq.~\eqref{eq:ULpoldamping}.
  Interestingly enough, one can obtain the eigenmodes of the upper and lower polariton branches while explicitly accounting for excitonic and photonic damping by writing the coupled damped oscillator equation~\cite{Kavokin2007},

  \begin{equation}
	  \hbar^2 V^2 = \left( E_{ex} - E - i \hbar \gamma_{ex} \right)\left( E_{ph} - E - i \hbar \gamma_{ph} \right),
	  \label{eq:coupledosc}
  \end{equation}
  where the two solutions of $E$ correspond exactly to the expressions given in Eq.~\eqref{eq:ULpoldamping}.

  According to Eq.~\eqref{eq:ULpoldamping}, the observable difference between the upper and lower polariton eigenergies at $\mathbf{k} = 0$, known as the Rabi splitting, is given by:

  \begin{equation}
	\hbar \Omega_R = E_{\text{UP}} - E_{\text{LP}} = 2 \sqrt{\hbar^2 V^2 + \frac{1}{4} \left[ \Delta E + i \hbar \left( \gamma_{ex} - \gamma_{ph} \right) \right]^2}.
	\label{eq:rabisplit}
  \end{equation}
  Let us focus on the case of zero detuning, $\Delta E = 0$.
  Then Eq.~\eqref{eq:rabisplit} reduces to:

  \begin{equation}
	\hbar \Omega_R = 2 \hbar \sqrt{V^2 - \left( \frac{\gamma_{ex} - \gamma_{ph}}{2} \right)^2}.
	\label{eq:rabisplitzdt}
  \end{equation}

  Now, if the exciton-photon coupling constant $V > |\gamma_{ex} - \gamma_{ph}|/2$, the system is said to be in the strong-coupling regime, where the UP/LP eigenenergies differ from the bare exciton and photon energies, and therefore the Rabi splitting, $\hbar \Omega_R$, is positive and real.
  If instead $V < |\gamma_{ex} - \gamma_{ph}|/2$, the system is in the weak-coupling regime, where $E_{\text{UP}}$ and $E_{\text{LP}}$ correspond to the exciton and photon energies, and the quantity $\hbar \Omega_R$ is imaginary.
  If $\gamma_{ex}$ and $\gamma_{ph}$ are much smaller than $V$, the Rabi splitting can be approximated by $\hbar \Omega_R \approx 2 \hbar V$.
  By calcuating $\hbar \Omega_R^0 \equiv 2 \hbar V$ and comparing it to $\hbar \Omega_R$, we can compare the strength of the exciton-photon interaction to the splitting between the upper and lower polariton eigenmodes, and in the process gain insight into the effect of the excitonic and photonic damping on the formation and properties of polaritons.

  Before presenting calculations of $\hbar \Omega_R^0$ and $\hbar \Omega_R$, let us discuss the choice of $\gamma_{ex}$ to be used in our calculations.
  Experimental studies of the excitonic properties in the FS Xenes are non-existent, and the sensitivity of the material parameters of silicene to the choice of substrate further complicates the generalization of experimental data for Silicene between substrates.
  On the other hand, the excitonic and optical properties of the TMDCs have been extensively studied for nearly a decade~\cite{Mak2010}, and TMDCs have been studied experimentally in optical microcavities for the past 5 years~\cite{Liu2014c,Dufferwiel2014,Schwarz2014,Dufferwiel2015}.
  The linewidth of the excitonic transition in the TMDCs has been observed to be roughly 11 meV at cryogenic temperatures~\cite{Dufferwiel2015} and appoximately $30$ meV at room-temperature~\cite{Liu2014c,Flatten2016,Wang2016,Lundt2017}.
  Since we consider the polaritonic properties at room-temperature in this work, we assume $\gamma_{ex} = (30~\text{meV})/\hbar \approx 5 \times 10^{13}$ s$^{-1}$ as an upper limit of $\gamma_{ex}$ to be used in calculations for both the FS Xenes and Type I/II Si.
  However, it is well documented that encapsulating TMDCs with $h$-BN strongly suppresses the excitonic linewidth, such that cryogenic linewidths are observed even at room temperature~\cite{Cadiz2017,Robert2017,Horng2018}.
  Therefore, we also consider a lower limit of $\gamma_{ex} = 10^{13}$ s$^{-1}$ for Type I/II Si.

  \begin{figure}[h]
	\centering
	\includegraphics[width=0.60\columnwidth]{1126-generic-5legend-eps-converted-to.pdf}\\
	\includegraphics[width=0.49\columnwidth]{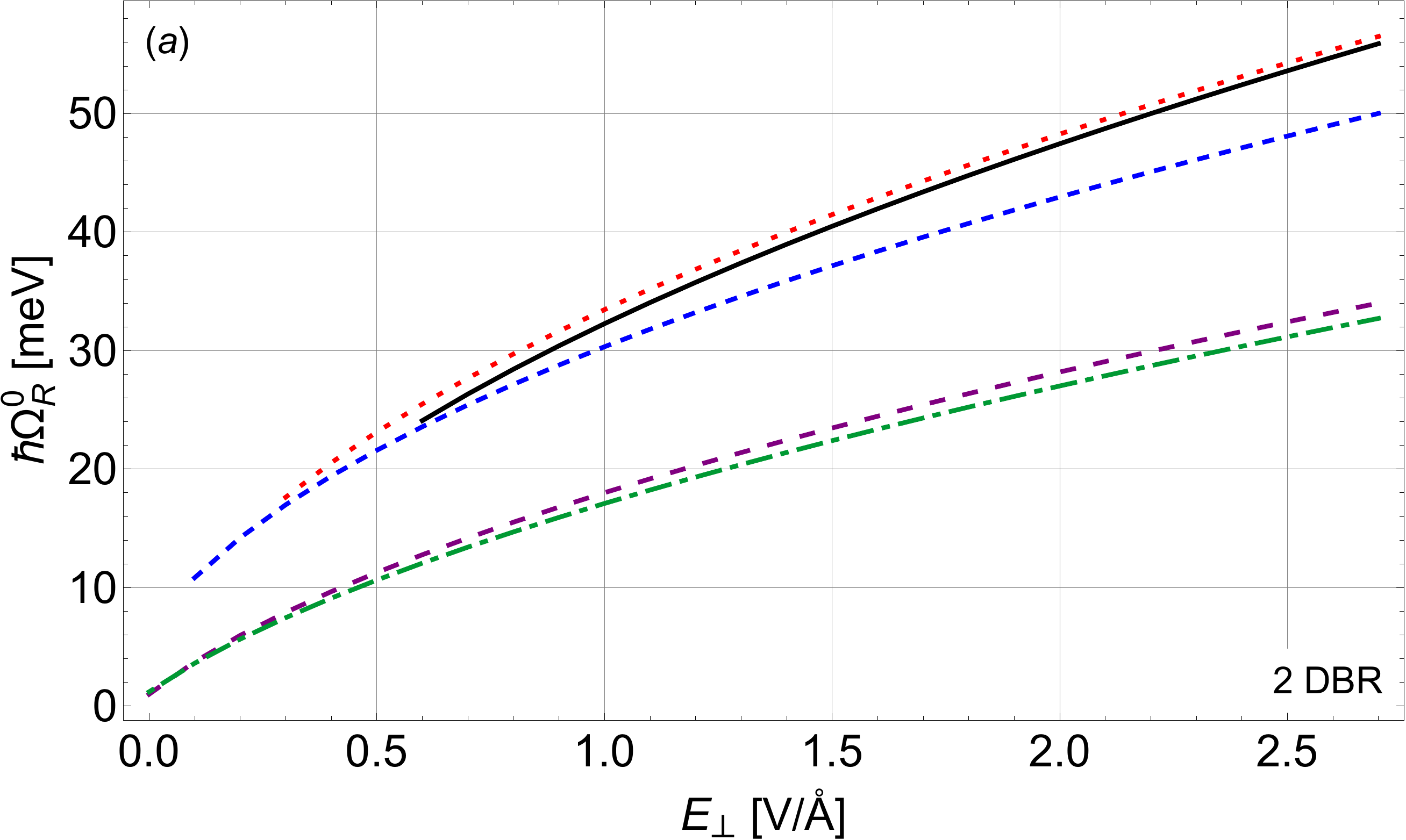}
	\includegraphics[width=0.49\columnwidth]{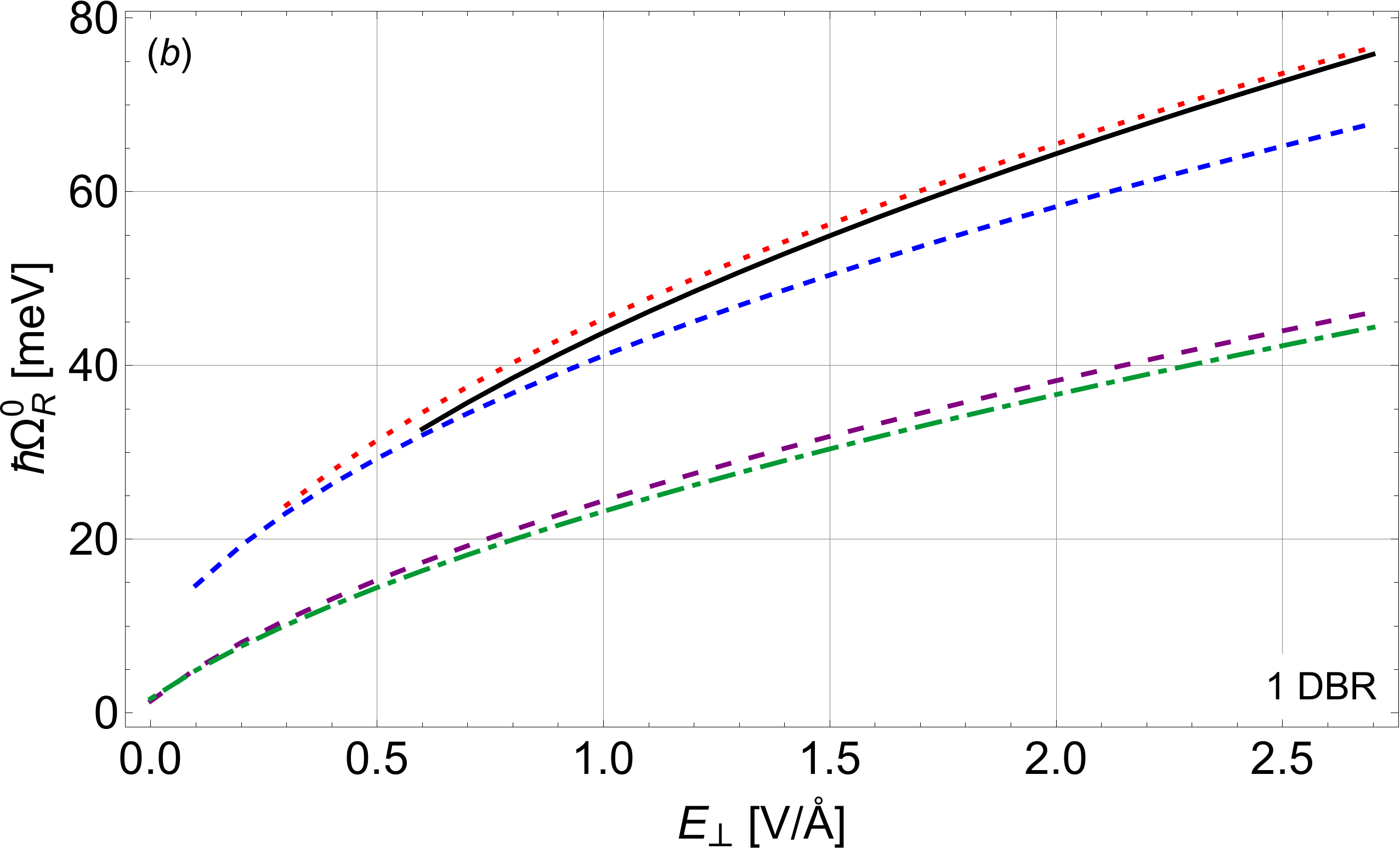}
	\includegraphics[width=0.49\columnwidth]{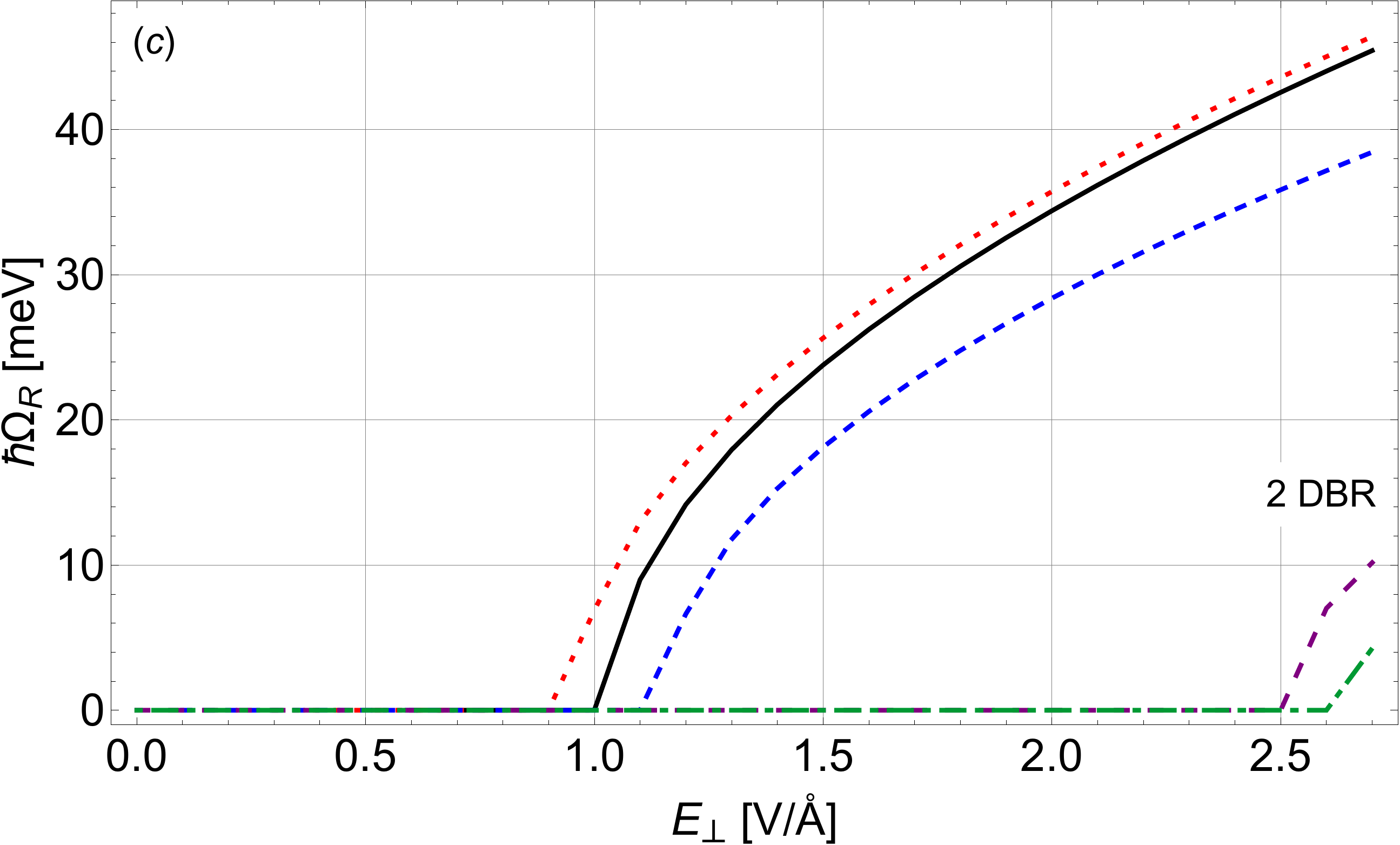}
	\includegraphics[width=0.49\columnwidth]{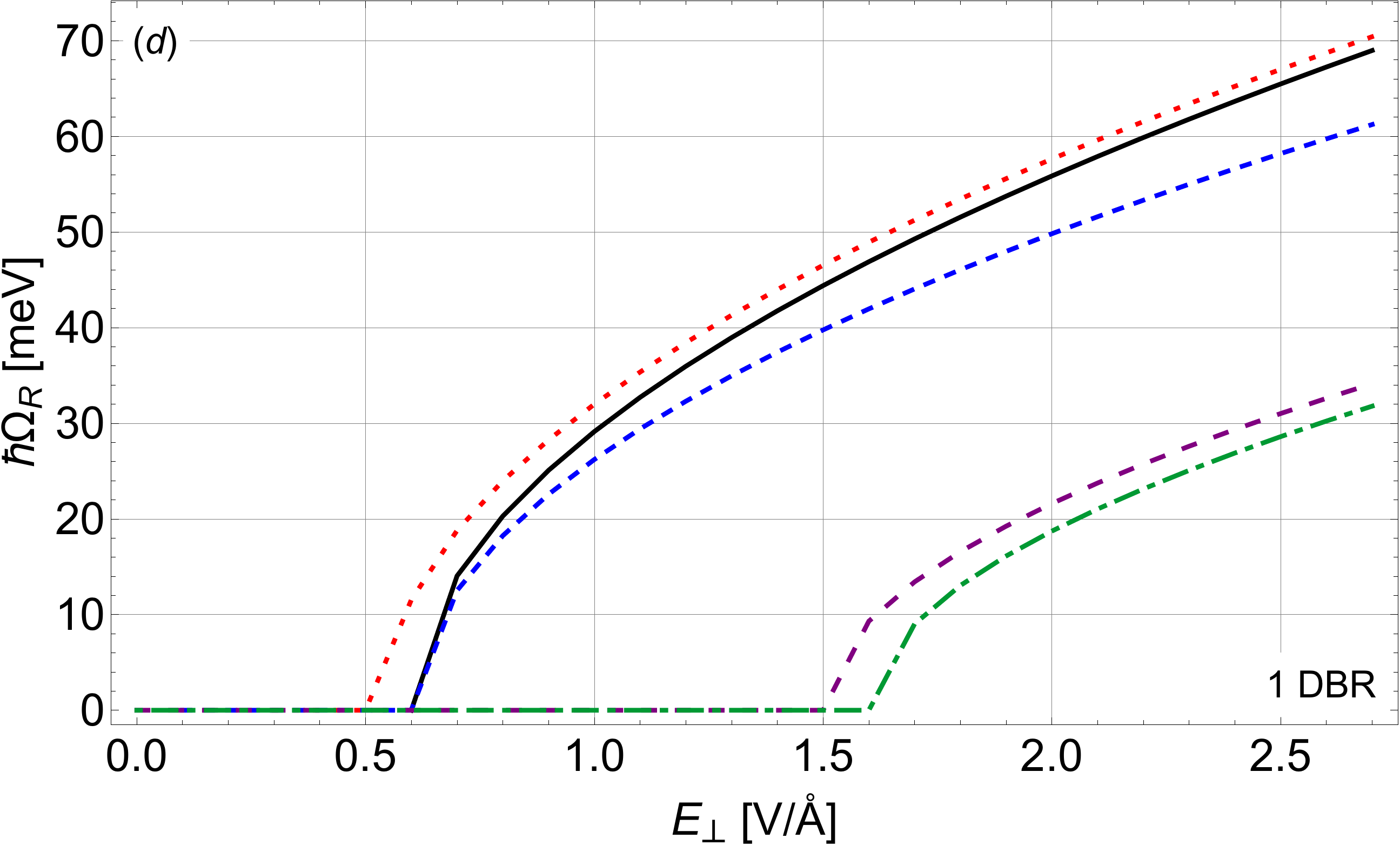}
	\caption{%
	  Dependence of the exciton-photon coupling constant, $\hbar \Omega_R^0$, on the external electric field, $E_\perp$, in each of the five materials, for $A$ excitons, in (a) a 2 DBR microcavity, and (b) a 1 DBR microcavity.
	  Dependence of the Rabi splitting with excitonic and photonic damping, $\hbar \Omega_R$, for $A$ excitons, in (c) a 2 DBR microcavity, and (d) a 1 DBR microcavity.
	}%
	\label{fig:rabiwwodamping}
  \end{figure}

  Comparisons of $\hbar \Omega_R^0$ and $\hbar \Omega_R$ across all five materials in both 2 DBR and 1 DBR microcavity designs are presented in Fig.~\ref{fig:rabiwwodamping}.
  Fig.~\ref{fig:rabiwwodamping}(a) shows the quantity $\hbar \Omega_R^0$ for $A$ excitons in each of the five materials in a 2 DBR microcavity, while Fig.~\ref{fig:rabiwwodamping}(b) shows the same quantity for a 1 DBR setup.
  Analysis of these results shows that the choice of microcavity configuration has a significant effect on the strength of the exciton-photon coupling constant $V$.
  Since $L_{\text{DBR}} \approx 4 L_c$, the extra factor of $L_{\text{DBR}}$ added to $L_{eff}$ in the 2 DBR configuration nearly doubles $L_{eff}$, and since $V \propto L_{eff}^{-1/2}$, we find that $V^{(\text{1 DBR})} \approx 1.35\times V^{(\text{2 DBR})}$ for all materials and at all electric fields.
  The factor of $\sqrt{\kappa}$ in the denominator of $V$ also significantly reduces $V$ in encapsulated Si ($\kappa = \varepsilon_{h\text{-BN}} = 4.89$), compared to the FS Xenes ($\kappa =1$).

  The large excitonic line broadening $\gamma_{ex}= 5\times10^{13}$ s$^{-1}$ and choice of $L_{eff}$ significantly affects the Rabi splitting, as presented in Figs.~\ref{fig:rabiwwodamping}(c) and (d).
  In the 2 DBR configuration shown in Fig.~\ref{fig:rabiwwodamping}(c), the FS Xenes do not enter the strong coupling regime until $E_\perp \approx 1.0$ V/\AA, while encapsulated Si is in the weak coupling regime until the electric field becomes extremely strong, $E_{\perp} \approx 2.5$ V/\AA.
  The 1 DBR configuration shown in Fig.~\ref{fig:rabiwwodamping}(d) demonstrates that the dependence of $L_{eff}$ on the cavity configuration has a significant effect on the onset of the strong coupling regime.
  In this microcavity setup, the FS Xenes enter the strong coupling regime around $E_\perp \approx 0.6$ V/\AA, while the transition to the strong coupling regime in encapsulated Si occurs around $E_\perp \approx 1.6$ V/\AA.
  In addition, the maximal value of $\hbar \Omega_R$ for each material is nearly twice as large in the 1 DBR case compared to the 2 DBR case.

  \begin{figure}[h]
	\centering
	\includegraphics[width=0.49\columnwidth]{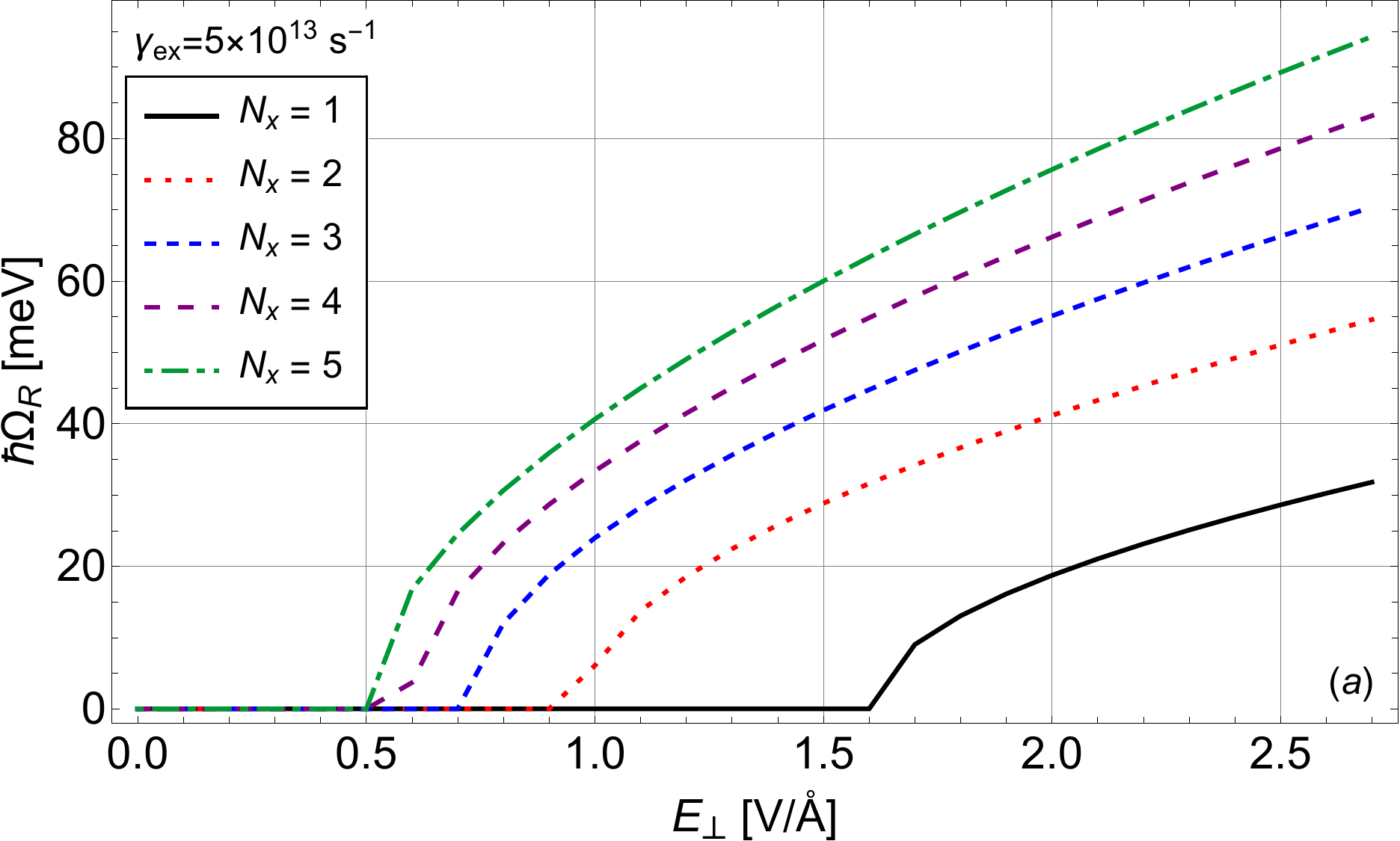}
	\includegraphics[width=0.49\columnwidth]{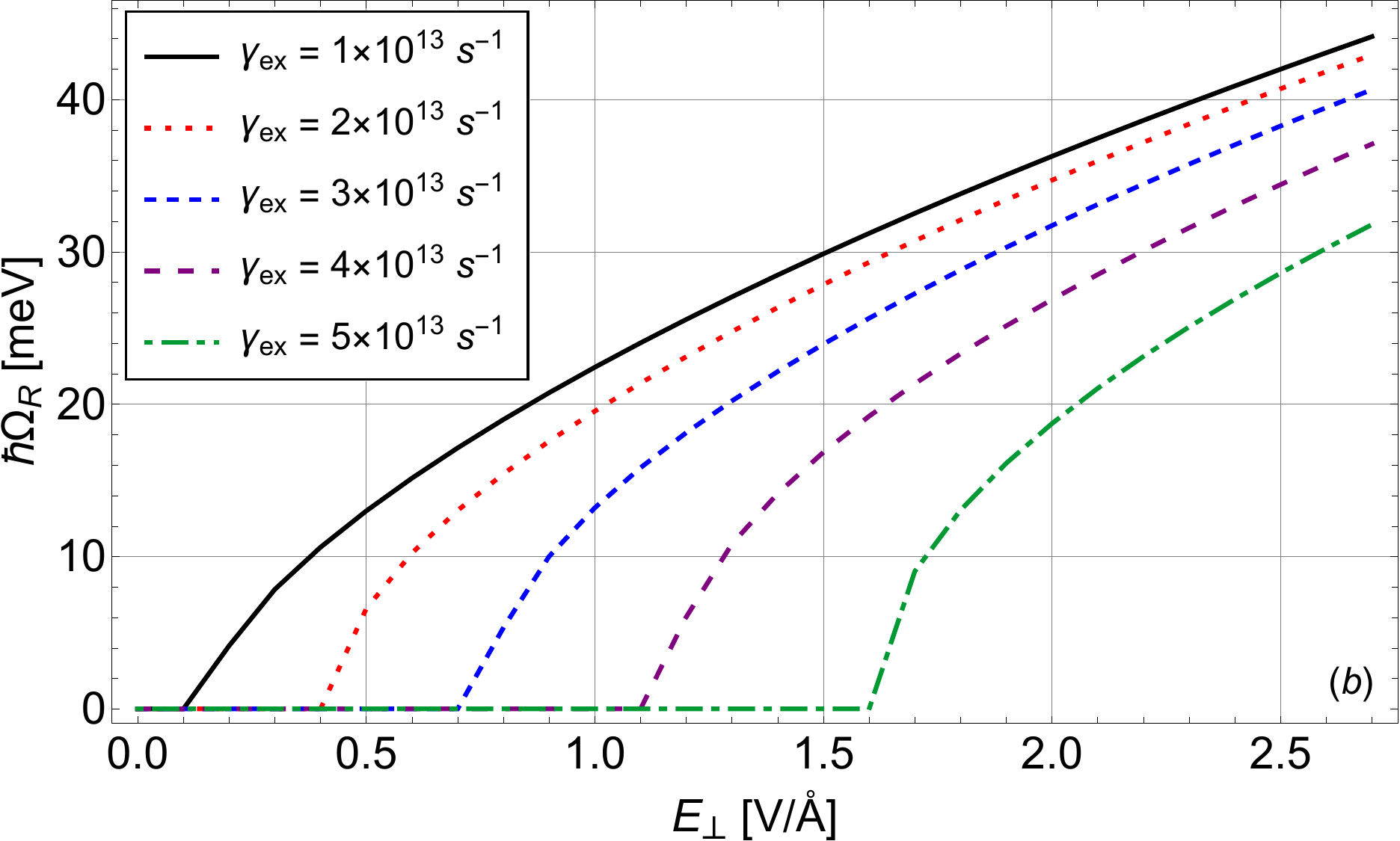}
	\caption{%
	  (a): Dependence of $\hbar \Omega_R$ on the electric field $E_\perp$ in Type II Si for different numbers of Si monolayers $N_X$ stacked on top of each other.
	  (b): Dependence of $\hbar \Omega_R$ on the electric field $E_\perp$ in Type II Si for different values of the exciton damping $\gamma_{ex}$.
	}%
	\label{fig:rabinxgamex}
  \end{figure}

  Next, we analyze the dependence of the Rabi splitting on the external electric field for different numbers of encapsulated Type II Si monolayers, $N_X$, stacked on top of each other, and on different values of the excitonic damping, $\gamma_{ex}$, shown in Fig.~\ref{fig:rabinxgamex}.
  In Fig.~\ref{fig:rabinxgamex}(a) we vary the number of Si monolayers in the microcavity, keeping $\gamma_{ex} = 5 \times 10^{13}$ s$^{-1}$.
  For $N_X > 1$, we consider a stack of Si monolayers, each separated by few-layer $h$-BN such that the Si monolayers do not interact with each other, while the height of the stack of $N_X$ Si monolayers with $h$-BN spacers remains negligible compared to $L_c$, so that $L_c$ and $L_{eff}$ do not need to be modified.
  We find that increasing $N_X$ to 3 brings the onset of the strong coupling regime in Type II Si to $E_{\perp} = 0.7$ V/\AA, and increases the maximum $\hbar \Omega_R$ at large electric fields to about 70 meV, which is quantitatively similar to $\hbar \Omega_R$ in the FS Xenes.

  In Fig.~\ref{fig:rabinxgamex}(b) we vary the exciton linewidth $\gamma_{ex}$ while keeping $N_X = 1$.
  We find that reducing $\gamma_{ex}$ reduces the value of $E_\perp$ at which the strong coupling regime is reached, but the maximal value of $\hbar \Omega_R$ at large electric fields is not increased as it is when $N_X$ is increased.
  This is because increasing $N_X$ increases $V$ itself, while reducing $\gamma_{ex}$ only causes $\hbar \Omega_R$ to converge towards $V$.

  Finally, let us comment on the relationship between the upper and lower polariton eigenenergies, the Rabi splitting, and the binding energy of polaritons, by which we mean the stability of polaritons against dissociation.
  At $\Delta E = 0$ and $\mathbf{k} = 0$, the lower (upper) polariton may dissociate into its constituent exciton and photon states if it gains (loses) an amount of energy equal to the difference between the lower (upper) polariton eigenenergy and the bare exciton/photon energy.
  Since the splitting of the upper/lower polariton eigenenergies is symmetric with respect to the (equal) bare exciton and photon energies, the binding energy of polaritons at $\Delta E = \mathbf{k} = 0$ is straightforwardly given by $E_{b,\text{UP/LP}} \equiv \lvert E_{\text{UP/LP}} - E_{ex/ph} \rvert = (\hbar \Omega_R)/2$.

  Since the splitting between the upper and lower polariton branches is symmetric with respect to the average of $E_{ex}$ and $E_{ph}$, in the case of either non-zero detuning or non-zero in-plane momentum, the polariton binding energy cannot be straightforwardly calculated as $(\hbar \Omega_R / 2)$.
  Restricting this example to the case where $E_{\text{UP}} (\mathbf{k}) > E_{ph} (\mathbf{k}) > E_{ex} (\mathbf{k}) > E_{\text{LP}} (\mathbf{k})$, we can see that upper polaritons would dissociate into photons if they lost energy equal to $E_{\text{UP}} - E_{ph}$, and lower polaritons would dissociate into excitons if they gained energy equal to $E_{ex} - E_{\text{LP}}$.
  Thus, the binding energy of polaritons can be given generally as

  \begin{align}
	  E_{b,\text{UP}}\left( \Delta E, \mathbf{k} \right) & = \lvert E_{\text{UP}} - \textit{Max}\left[ E_{ex},~E_{ph} \right] \rvert \nonumber \\
	  E_{b,\text{LP}}\left( \Delta E, \mathbf{k} \right) & = \lvert E_{\text{LP}} - \textit{Min}\left[ E_{ex},~E_{ph} \right] \rvert.
	\label{eq:ebuplp}
  \end{align}

\section{\label{sec:superfluidity}Tunable superfluidity of polaritons in an Xene monolayer in an open microcavity}

  In the previous section we considered a very dilute system of noninteracting polaritons when the exciton-exciton interaction term in the Hamiltonian~\eqref{eq:cavityham} is neglected.
  Now let us consider a weakly interacting Bose gas of polaritons, taking into account the exciton-exciton interaction.
  In this low-density limit, a system of excitons can be treated purely as bosons if one includes an interaction potential that accounts for the fermionic nature of the constituent electrons and holes~\cite{Ciuti1998,De-Leon2001}.
  For small wave vectors satisfying $\mathbf{q} \ll a_{2D}^{-1}$, where $a_{2D}$ is the 2D exciton Bohr radius, the exciton-exciton repulsion can be approximated by a contact potential, $U_{\mathbf{q}} \approx U_{0} \equiv 6 E_b a_{2D}^2$~\cite{Kavokin2003,Laussy2010,Vasilevskiy2015}.

  Diagonalizing Eq.~\eqref{eq:cavityham} using the same procedure as before without discarding the exciton-exciton interaction term, we obtain the Hamiltonian for a system of interacting lower polaritons:

  \begin{equation}
	  \hat{H}_{\text{LP}} = \sum_\mathbf{k} E_{\text{LP}} (\mathbf{k}) \hat{p}_\mathbf{P}^\dag \hat{p}_\mathbf{P} +
	  \frac{1}{2 A} \sum_{\mathbf{k},\mathbf{k}',\mathbf{q}} U_{\mathbf{k},\mathbf{k'},\mathbf{q}} \hat{p}_{\mathbf{k}+\mathbf{q}}^\dag \hat{p}_{\mathbf{k}'-\mathbf{q}}^\dag \hat{p}_\mathbf{k} \hat{p}_{\mathbf{k}'},
	\label{eq:LPham}
  \end{equation}
  where the second term now describes the repulsive polariton-polariton interaction potential, corresponding to the fourth term in Eq.~\eqref{eq:cavityham}.
  Eq.~\eqref{eq:LPham} therefore corresponds to a dilute, weakly interacting Bose gas of lower polaritons.
  Since polaritons interact entirely via their excitonic component, the polariton-polariton interaction potential must be proportional to the exciton-exciton interaction, $U_{\mathbf{k},\mathbf{k'},\mathbf{q}} \propto U_{\mathbf{q}}$, and is given by~\cite{Kavokin2003,Laussy2010,Vasilevskiy2015}:
  \begin{equation}
	U_{\mathbf{k},\mathbf{k'},\mathbf{q}} = 6 E_b a_{2D}^2 X_{\mathbf{k+q}} X_{\mathbf{k'}} X_{\mathbf{k'-q}} X_{\mathbf{k}}.
	\label{eq:Ueff}
  \end{equation}

  In the Bogoliubov approximation, the sound spectrum of collective excitations at low momenta in a dilute, weakly interacting Bose gas is given by $\epsilon (\mathbf{k}) = c_s \mathbf{k}$, where $c_s$ is the sound velocity~\cite{Abrikosovbook,Griffinbook}:

  \begin{equation}
	  c_s = \sqrt{\frac{U_{eff} n_{\text{LP}}}{M_{\text{LP}}}}.
	\label{eq:bogocs}
  \end{equation}
  In Eq.~\eqref{eq:bogocs}, $U_{eff} \equiv U_{(\mathbf{k},\mathbf{k}',\mathbf{q}) \approx 0} = 6 E_b a_{2D}^{2} \lvert X \rvert^4$ is the effective polariton-polariton interaction potential in the limit of small momenta, $n_{\text{LP}}$ is the 2D concentration of lower polaritons, and $M_{\text{LP}}$ is the effective mass of lower polaritons, given by:

  \begin{equation}
	M_{\text{LP}}^{-1} \left( \mathbf{k} \right) = \lvert X \rvert^2 M_{ex}^{-1} + \lvert C \rvert^2 m_{ph}^{-1}.
	\label{eq:mpol}
  \end{equation}
  The Hopfield coefficients $\lvert X \rvert$ and $\lvert C \rvert$ in both Eq.~\eqref{eq:mpol} and in $U_{eff}$ are evaluated in the limit $\mathbf{k} \to 0$, but we note that their values here still depend on the detuning $\Delta E$ between the exciton and photon eigenenergies.

  A dilute 2D gas of weakly interacting bosons experiences a BKT transition to the superfluid phase at a critical temperature~\cite{Berezinskii1971,Berezinskii1972,Kosterlitz1972,Kosterlitz1973,Nelson1977}:

  \begin{equation}
	  T_{c} = \frac{\pi \hbar^2 n_s(T_{c})}{2 k_B M_{\text{LP}}}.
	\label{eq:KTTc}
  \end{equation}
  In Eq.~\eqref{eq:KTTc}, $k_B$ is the Boltzmann constant and $n_s(T) = n_{\text{LP}} - n_n (T)$ is the superfluid concentration at temperature $T$, where $n_n$ is the 2D concentration of the normal component of the polariton Bose fluid, given by~\cite{Abrikosovbook}:

  \begin{equation}
	  n_s(T) = n_{\text{LP}} - \frac{3 \zeta(3)}{2 \pi \hbar^2} \frac{s k_B^3 T^3}{c_s^4 M_{\text{LP}}},
	\label{eq:nsdensity}
  \end{equation}
  where the spin degeneracy factor $s=1$.
  Setting $n_s(T) = 0$ one obtains the critical temperature in the mean-field approximation, or in other words, the temperature at which the local concentration of the superfluid component, $n_s (T)$, vanishes:

  \begin{equation}
	T_c^0 = \left[ \frac{2 \pi \hbar^2 U_{eff}^2}{3 \zeta(3) M_{\text{LP}}} \right]^{1/3} \frac{n_{\text{LP}}}{k_B}.
	  \label{eq:tc0}
  \end{equation}
  Solving Eq.~\eqref{eq:tc0} for $n_{\text{LP}}$, one effectively obtains the maximum 2D concentration of the normal component of the 2D Bose gas of LP at a given temperature $T$:
  \begin{equation}
	n_c^0 = \left[ \frac{3 \zeta(3) M_{\text{LP}}}{2 \pi \hbar^2 U_{eff}^2} \right]^{1/3} k_B T_c^0.
	  \label{eq:critdens}
  \end{equation}
  Therefore, $n_c^0$ is the critical LP concentration in the mean-field approximation, for a given $T$.
  In other words, a 2D weakly interacting Bose gas of LP can only sustain a finite LP concentration in the normal phase; as more LP are added they occupy the degenerate superfluid state.

  Substituting Eq.~\eqref{eq:nsdensity} into Eq.~\eqref{eq:KTTc}, one obtains a cubic equation for the BKT transition temperature, which has the following solution:

  \begin{equation}
	T_c = \left[ \left( 1 + \sqrt{\frac{32}{27} \left( \frac{M_{\text{LP}} k_B T_c^0}{\pi \hbar^2 n_{\text{LP}}} \right)^{3} + 1 } \right)^{1/3} + \left( 1 - \sqrt{\frac{32}{27} \left( \frac{M_{\text{LP}} k_B T_c^0}{\pi \hbar^2 n_{\text{LP}}} \right)^{3} + 1 } \right)^{1/3} \right] \frac{T_c^0}{2^{1/3}}.
	\label{eq:tcfull}
  \end{equation}

\begin{figure}[h]
	\centering
	\includegraphics[width=0.60\columnwidth]{1126-generic-5legend-eps-converted-to.pdf}
	\includegraphics[width=0.49\columnwidth]{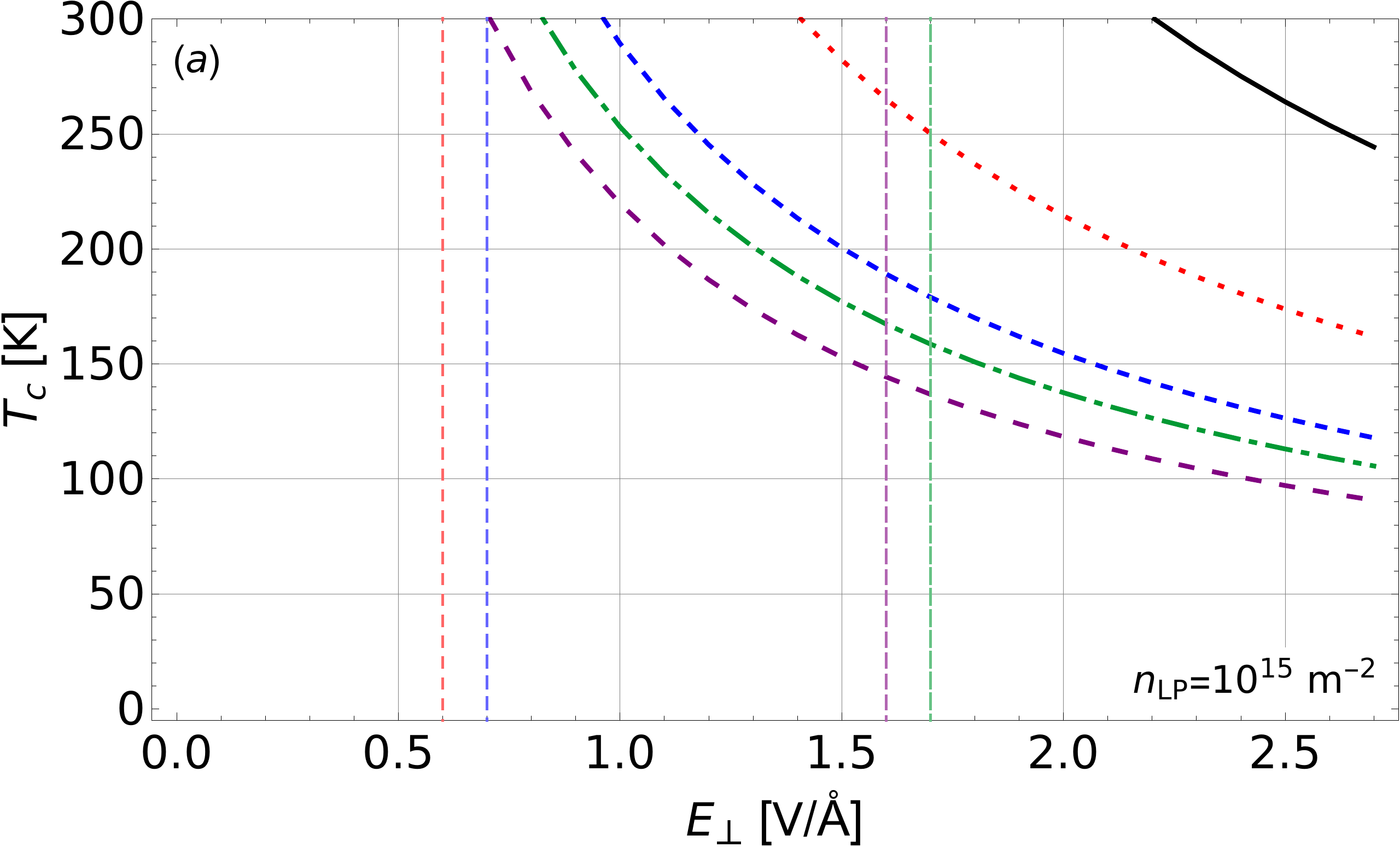}
	\includegraphics[width=0.49\columnwidth]{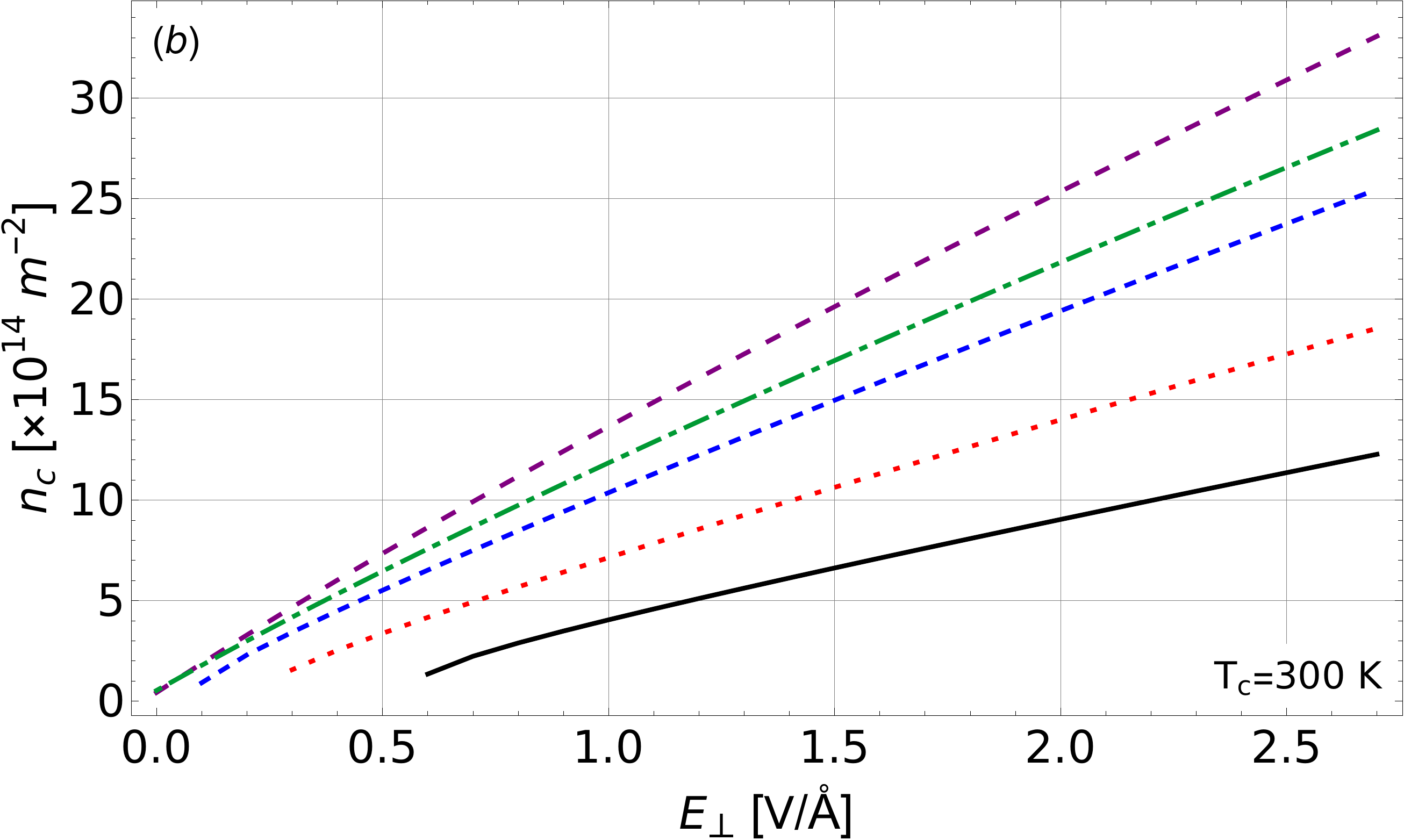}
	\caption{\label{fig:simpletcline}%
	  (a): Dependence of $T_{c}$ on $E_{\perp}$.
	  The vertical dashed lines denote the minimum $E_\perp$ in which the corresponding material is in the strong coupling regime in a 1 DBR configuration based on Fig.~\ref{fig:rabiwwodamping}.
	  (b): Dependence of the critical LP concentration $n_{c}$ on the external electric field, for $A$ polaritons, at $T_{c} = 300$ K.
	}%
  \end{figure}

  Analysis of Eqs.~\eqref{eq:deltaez},~\eqref{eq:effmassEz},~\eqref{eq:mpol}, and~\eqref{eq:tcfull} shows that the BKT transition temperature $T_c$ depends on the polariton concentration, $n_{\text{LP}}$, the applied external electric field, $E_{\perp}$, and on the properties of the microcavity.
  The results of calculations of the dependence of the BKT transition temperature and the critical LP concentration in the mean-field approximation on the external electric field are presented in Fig.~\ref{fig:simpletcline}.

   Using Eq.~\eqref{eq:tcfull}, we calculated the dependence of the BKT critical temperature $T_c$ on the external electric field, $E_{\perp}$, for each Xene at $n_{\text{LP}}=10^{15}$ m$^{-2}$, shown in Fig.~\ref{fig:simpletcline}(a).
  Analysis of the figure indicates that $T_{c}$ decreases as the external electric field, $E_{\perp}$, is increased, and that FS Si has by far the largest $T_{c}$ at all values of $E_{\perp}$, while Type I and Type II encapsulated Si have the smallest $T_{c}$.
  Also shown in Fig.~\ref{fig:simpletcline}(a) are vertical dashed lines which denote the $E_{\perp}$ at which the strong coupling regime is reached for the corresponding material in a 1 DBR microcavity configuration.
  While the FS Xenes are in the strong coupling regime when the corresponding $T_{c} > 300$ K, Type I/II encapsulated Si are not in the strong coupling regime when $T_{c} > 300$ K.

  Calculations of the critical LP concentration, $n_{\text{LP}}$, for the BKT transition to the superfluid phase for fixed critical temperature $T_c = 300$ K as a function of $E_{\perp}$ are presented in Fig.~\ref{fig:simpletcline}(b).
  One can recover the superfluid phase in encapsulated Si at high $E_{\perp}$ (when encapsulated Si is in the strong coupling regime) by increasing the LP concentration to about $n_{\text{LP}} = 2 \times 10^{15}$ m$^{-2}$, an increase of only a factor of two compared to the concentration used to calculate the results in Fig.~\ref{fig:simpletcline}(a), $n_{\text{LP}} = 10^{15}$ m$^{-2}$.
  Therefore, the nearly linear relationship between $n_{c}$ and $E_{\perp}$ at fixed $T$ means that increasing $n_{\text{LP}}$ by a factor of two in turn increases $T_{c}$ by a factor of two at a given $E_{\perp}$.

  Another interesting aspect of the BKT phase transition in polaritons is the relationship between $T_{c}$ and the detuning, $\Delta E$.
  Since both the effective polariton-polariton interaction potential, $U_{eff}$, and the LP mass, $M_{\text{LP}}$, depend on the Hopfield coefficients $\lvert X \rvert^2$ and $\lvert C \rvert^2$, which in turn depend on the detuning, $\Delta E$, the BKT critical temperature therefore depends non-monotonically on the detuning.
  For positive $\Delta E$, the LP becomes more exciton-like, increasing the strength of the effective interaction potential, which increases $T_{c}$, but at the same time, $M_{\text{LP}}$ also increases, which decreases $T_{c}$.
  Since $U_{eff}$ can only increase by a factor of four, while $M_{\text{LP}}$ can vary by several orders of magnitude between $m_{ph}$ and $M_{ex}$, we find that a local maximum in $T_{c}$ is reached for small positive values of $\Delta E$.
  Specifically, in FS Si, for small $E_{\perp}$, the value of $\Delta E$ which maximizes $T_{c}$ corresponds to $E_{ph} \approx 1.07 E_{ex}$, and as $E_{\perp}$ increases, the maximizing value of $\Delta E$ approaches $E_{ph} \approx 1.03 E_{ex}$, or in other words, the percent-detuning which maximizes $T_c$ in FS Si lies between 4-7\%, depending on $E_{\perp}$.
  Interestingly, in all materials except FS Si, the percent-detuning which maximizes $T_c$ lies between approximately 1-4\%.
  Furthermore, the difference in $T_{c}$ between the maximal detuning and zero detuning is about 35\% for all values of $E_\perp$ in all materials, quite a significant increase.

  However, in Sec.~\ref{sec:polaritons} it was mentioned that any non-zero detuning reduces the LP binding energy by reducing the energy difference between $E_{\text{LP}}$ and the lesser of $E_{ex}$ or $E_{ph}$.
  Our calculations show that as the detuning $\Delta E > 0$ is increased, the LP binding energy decreases faster than $T_{c}$ increases, making detuning an ineffective mechanism for maximizing $T_{c}$ and $E_{b,\text{LP}}$ simultaneously.
  On the other hand, for $N_X > 1$ in encapsulated Si, varying the detuning may be an effective way to increase $T_{c}$, since the enhanced Rabi splitting compared to $N_X = 1$ can allow for LP which are still stable at $T_{c}$ despite the reduction in $E_{b,\text{LP}}$ for non-zero detunings.
  At temperatures $T$ where $E_{b,\text{LP}} \gg k_B T$, changing the detuning can still be used to tune $T_{c}$, but examination of Fig.~\ref{fig:simpletcline} indicates that one should encounter few difficulties achieving a stable superfluid of LP at relatively low temperatures in any case.

  In essence, simulatenously maximizing $E_{b,\text{LP}}$ and $T_{c}$ subject to a variety of constraints on material and environmental parameters such as $n_{\text{LP}}$, $\Delta E$, $N_X$, and $\gamma_{ex}$ is a classic example of an optimization problem, which we analyze in detail in the next Section.

  \section{\label{sec:opt}Analysis of the Rabi splitting and superfluid critical temperature}

  Earlier, we calculated the dependence of the Rabi splitting on $E_{\perp}$, $\gamma_{ex}$, and $N_X$, and found that polaritons in the Xenes should be stable at room-temperature for some combination of large $E_{\perp}$ and small $\gamma_{ex}$, and that increasing $N_X$ is a straightforward way of significantly increasing $\hbar \Omega_R$ in encapsulated Si.
  Then, calculations of Eq.~\eqref{eq:tcfull} revealed that $T_{c}$ in all materials is extremely high at small $E_{\perp}$, and even for $E_{\perp} > 2.0$ V/\AA, $T_{c}$ exceeds 100 K.
  Let us now analyze the intersection of the ``high Rabi splitting'' and ``high $T_{c}$'' regimes and the dependence of these regimes on experimental parameters such as $n_{\text{LP}}$, $\Delta E$, $\gamma_{ex}$, and $N_X$.
  We primarily focus on encapsulated (Type II) Si since such a setup is already very similar to previous experimental work on polaritons in TMDCs encapsulated by $h$-BN~\cite{Liu2014c,Vasilevskiy2015,Flatten2016}.
  Instead of addressing each of the FS Xenes individually, we focus on FS Si because it shows by far the largest $T_{c}$ of the FS Xenes, while the FS Xenes all have similar $\hbar \Omega_R$.

  At a minimum, three conditions must be met in order to obtain a stable LP superfluid at room temperature: (i) the exciton binding energy must exceed 26 meV, (ii) the LP binding energy, given by $\hbar \Omega_R /2$ at $\Delta E = 0$ and by $E_{ex} - E_{\text{LP}}$ at $\Delta E > 0$, must exceed 26 meV, and (iii) the critical temperature for the BKT superfluid phase transition must be at least 300 K.
  Since the exciton binding energy in all materials exceeds 100 meV for $E_{\perp} > 0.5$ V/\AA, the excitons are certainly stable in the range of $E_{\perp}$ considered~\cite{Brunetti2018b}.
  The conditions therefore reduce to: $E_{b,\text{LP}} > k_B T$ and $T_{c} > T$.

\begin{figure}[t]
	\centering
	\includegraphics[width=0.49\columnwidth]{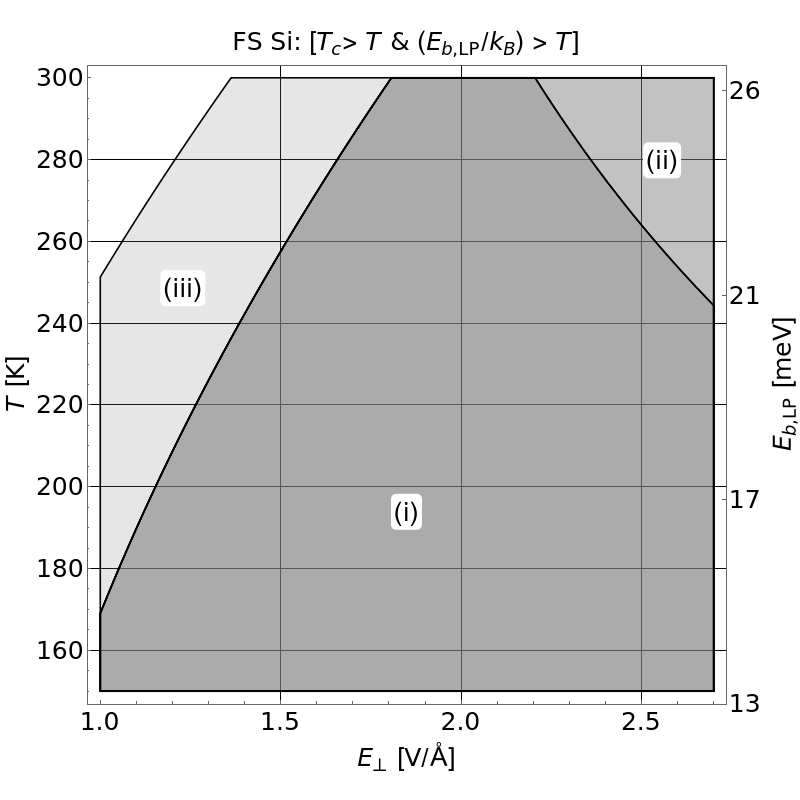}
	\caption{%
	  Shaded regions show where $T_{c}$ and $E_{b,\text{LP}}$ both exceed $T$ (along vertical axis) in FS Si.
	  The labeled regions denote the following combinations of parameters: (i) $n_{\text{LP}} = 10^{15}$ m$^{-2}$; (ii) $n_{\text{LP}} = 1.25 \times 10^{15}$ m$^{-2}$; (iii) $n_{\text{LP}} = 10^{15}$ m$^{-2}$; for (i) and (ii), $\gamma_{ex} = 5 \times 10^{13}$ s$^{-1}$, in (iii), $\gamma_{ex} = 10^{13}$ s$^{-1}$.
	  Regions (i) and (ii) share the same left-hand-side boundary (same $\gamma_{ex}$), while regions (i) and (iii) share the same right-hand-side-boundary (same $n_{\text{LP}}$).
	}
	\label{fig:sirpvsept}
  \end{figure}

\begin{figure}[h]
	\centering
	\includegraphics[width=0.49\columnwidth]{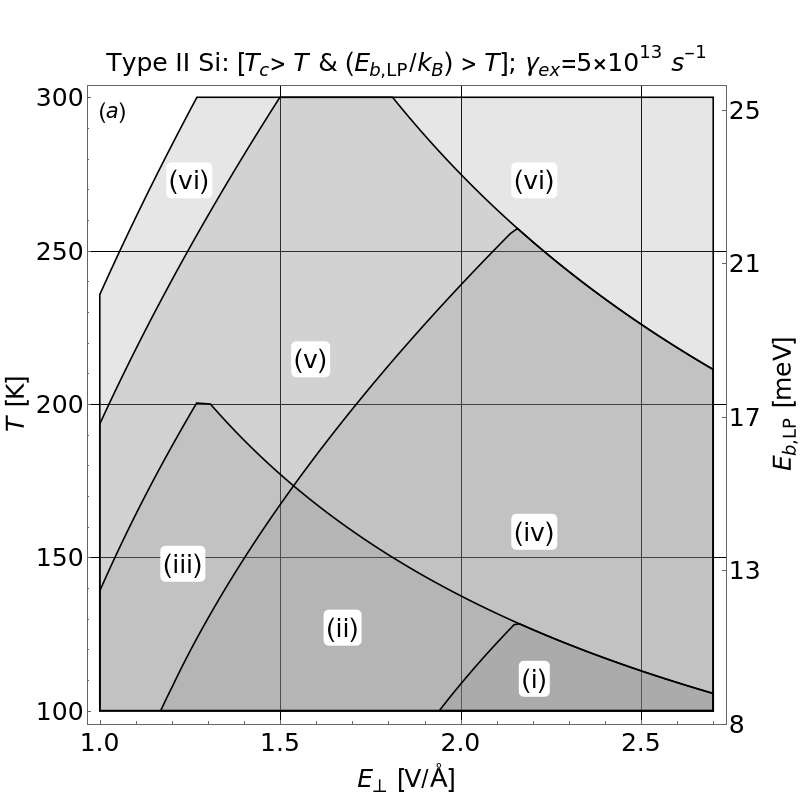}
	\includegraphics[width=0.49\columnwidth]{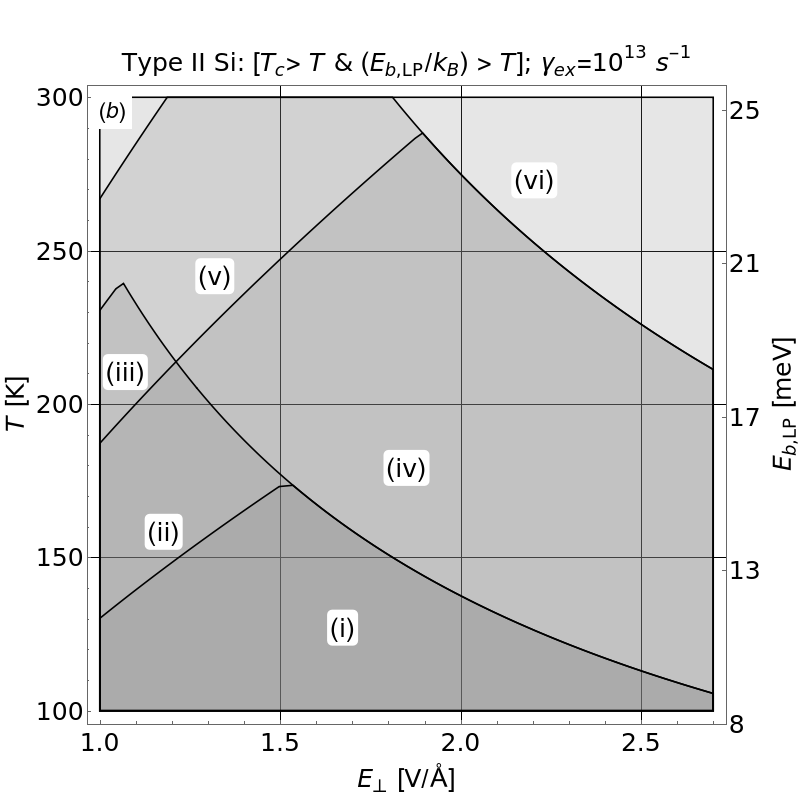}
	\caption{\label{fig:t2rpvsept}%
	  (a): Shaded regions denote ranges of $E_{\perp}$ and $T$ where a stable LP superfluid is expected to form for the combination of material parameters specified by the region labels.
	  (i),(ii),(iii): $n_{\text{LP}} = 10^{15}$ m$^{-2}$ and $N_X = 1, 2, 3$, respectively; (iv),(v): $n_{\text{LP}} = 2 \times 10^{15}$ m$^{-2}$ and $N_X = 2,4$, respectively; (vi): $n_{\text{LP}} = 3 \times 10^{15}$ m$^{-2}$ and $N_X = 5$.
	  (b): Shaded regions correspond to the same parameters as in (a), but with $\gamma_{ex} = 10^{13}$ s$^{-1}$.
	}
  \end{figure}

  Figs.~\ref{fig:sirpvsept} and~\ref{fig:t2rpvsept} depict regions where both $T_{c}$ and $E_{b,\text{LP}}/k_B$ exceed $T$ (along the vertical axis) for the range of $1.0 < E_{\perp} < 2.7$ V/\AA~in FS Si or Type II encapsulated Si.
  Each region therefore represents the range of external electric field, $E_{\perp}$, and ambient temperature, $T$, for which a stable LP superfluid could form, based on the choice of material parameters denoted by the lower case roman numeral in each region.
  The left-hand boundary of each region is formed by the curve $E_{b,\text{LP}}(E_{\perp}) = k_B T$, while the right-hand boundary of each region is formed by the curve $T_{c}(E_{\perp}) = T$.

  Fig.~\ref{fig:sirpvsept} shows that FS Si should support a stable LP superfluid at $T = 300$ K under the following conditions: (i) $E_{\perp} \in [1.8,2.2]$ V/\AA; (ii) $E_{\perp} > 1.8$ V/\AA; (iii) $E_{\perp} \in [1.3,2.2]$ V/\AA.
  In other words, for $\gamma_{ex} = 5 \times 10^{13}$ s$^{-1}$ and $n_{\text{LP}} = 10^{15}$ m$^{-2}$, a stable LP superfluid should form at $T = 300$ K for $1.8~\text{V/\AA} < E_{\perp} < 2.2~\text{V/\AA}$; if $n_{\text{LP}} \geq 1.25 \times 10^{15}$ m$^{-2}$, a stable superfluid should form for all $E_{\perp} > 1.8$ V/\AA.

  Fig.~\ref{fig:t2rpvsept}(a) shows the same information as in Fig.~\ref{fig:sirpvsept} but for Type II Si with $\gamma_{ex} = 5 \times 10^{13}$ s$^{-1}$.
  The different regions correspond to the overlap of $T_{c} > T$ and $E_{b,\text{LP}}/k_B > T$ for different combinations of $N_X$ and $n_{\text{LP}}$.
  As was shown in Fig.~\ref{fig:simpletcline}, $T_{c} < 300$ K in Type II Si for $n_{\text{LP}} = 10^{15}$ m$^{-2}$, so that the prospect of room-temperature superfluidity is only feasible for $n_{\text{LP}} > 2 \times 10^{15}$ m$^{-2}$ and $N_X > 4$ when $E_{\perp} \in [1.5, 1.8]$ V/\AA.
  If instead $n_{\text{LP}} = 3 \times 10^{15}$ m$^{-2}$, by following the curves of the region boundaries we find that a stable room-temperature superfluid could be achieved for $N_X = 3$ for $E_{\perp} \gtrapprox 2.0$ V/\AA.
  At $n_{\text{LP}} = 3 \times 10^{15}$ m$^{-2}$ and $N_X = 4$, a stable room-temperature superfluid could be expected to form for $E_{\perp} \gtrapprox 1.5$ V/\AA.

  In Fig.~\ref{fig:t2rpvsept}(b) we consider the conditions for a stable LP superfluid in Type II Si with $\gamma_{ex} = 10^{13}$ s$^{-1}$, while the labeled regions correspond to the same combinations of material parameters as in Fig.~\ref{fig:t2rpvsept}.
  The benefits of the possible reduction in $\gamma_{ex}$ due to $h$-BN encapsulation are clearly displayed here \textendash{} the increase in $E_{b,\text{LP}}$ means that at $T = 300$ K a stable LP superfluid could form at $n_{\text{LP}} = 2 \times 10^{15}$ m$^{-2}$ for $1.2$ V/\AA~$< E_{\perp} < 1.8$ V/\AA~with $N_X = 4$, while for $N_X = 3$ the corresponding range of $E_{\perp}$ is roughly 1.4 - 1.8 V/\AA.
  These ranges of $E_{\perp}$ are notably smaller than the corresponding range of $E_{\perp}$ shown in Fig.~\ref{fig:sirpvsept} for FS Si, but the values of $E_{\perp}$ are themselves also smaller than in FS Si.

  Finally, let us address the effect that non-zero detuning has on the existence of a stable LP superfluid.
  As mentioned in Sec.~\ref{sec:superfluidity}, $T_{c}$ increases if the photon energy is slightly blueshifted from the exciton energy, and $T_{c}$ reaches a maximum when the photon energy exceeds the exciton energy by approximately 1-2\% (the precise value of the percentage-detuning which maximizes $T_{c}$ depends on $E_{\perp}$ but does not depend on $n_{\text{LP}}$ \textendash{} the maximal value of the percent-detuning decreases as $E_{\perp}$ increases).
  When the detuning is increased from zero, two qualitative changes to the region of stability adversely affect the feasibility of high-temperature superfluidity.
  First, the temperature at which the curves $E_{b,\text{LP}}(E_{\perp})$ and $T_{c}(E_{\perp})$ intersect decreases.
  The intersection of these curves can be identified as the pointed apex of the regions in Figs.~\ref{fig:sirpvsept} and~\ref{fig:t2rpvsept}.
  Second, the range of $E_{\perp}$ which fall within a particular region widens, but the value of $E_{\perp}$ at the beginning and end of the range increases, or in other words, the region of $E_{\perp}$ widens but shifts towards greater $E_{\perp}$ overall.
  In general, we consider it advantageous to minimize the $E_{\perp}$ required to reach the SF regime since applying and maintaining a strong static electric field imposes additional challenges on device design and fabrication.

  \section{\label{sec:discussion}Discussion}

  We began by briefly reviewing the theoretical description of the electronic properties of Xenes in an external electric field and calculated the excitonic transition energy, $E_{ex} = 2 \Delta_{\text{SO}} - E_b$.
  The material with the smallest $E_{ex}$ is FS Si, which increases roughly linearly from $E_{ex} \approx 200$ meV at $E_{\perp} = 1.0$ V/\AA, to $E_{ex} \approx 1.6$ eV at $E_{\perp} = 2.7$ V/\AA.
  Despite the enhanced dielectric screening of encapsulated Si $(\kappa = 4.89)$ compared to FS Si ($\kappa = 1$), the modified material parameters of encapsulated Si, which are more similar to FS Ge than to FS Si, result in a maximal $E_{ex} \approx 2.2$ eV at $E_{\perp} = 2.7$ V/\AA.
  The maximum values for FS Ge and FS Sn are $E_{ex} \approx 3,~4.2$ eV, respectively.

  It was shown that the exciton-photon coupling strength, $\hbar \Omega_R^0$, can, at high electric fields, reach about $55$ meV in the FS Xenes and about $35$ meV in encapsulated Si for a 2 DBR microcavity configuration, while for a 1 DBR microcavity, the corresponding maximal values of $\hbar \Omega_R^0$ for the FS Xenes and encapsulated Si are about $80$ meV and $45$ meV, respectively.
  However, the large value of $\gamma_{ex}$ used here strongly suppresses the Rabi splitting $\hbar \Omega_R$, such that the polariton system is in the weak coupling regime until about $E_\perp = 1.0$ V/\AA~(0.5 V/\AA) in the FS Xenes and until about $E_{\perp} = 2.5$ V/\AA~(1.5 V/\AA) in encapsulated Si for 2 DBR (1 DBR) microcavity configurations.
  The significant reduction in $\hbar \Omega_R$ means that lower polaritons are never stable at room-temperature in a 2 DBR microcavity, while for a 1 DBR cavity, LP are only stable at room-temperature in the FS Xenes beyond about $E_{\perp} = 2.2$ V/\AA.
  Since $\gamma_{ex} = 5 \times 10^{13}$ s$^{-1}$ is borrowed from experimental determinations of the excitonic broadening in TMDCs at room temperature, it is possible that $\gamma_{ex}$ in the Xenes may be smaller, in which case we find that the strong coupling regime is reached at much lower $E_{\perp}$, though the maximal value of $\hbar \Omega_R$ at high $E_{\perp}$ does not change much.
  On the other hand, increasing $N_X$ appears to be a promising way to significantly increase the Rabi splitting at all values of $E_{\perp}$, since increasing $V$ depends directly on $N_X$.

  We then considered the conditions under which a weakly interacting Bose gas of lower polaritons would undergo a BKT phase transition to the superfluid phase.
  At typical LP concentrations $n_{\text{LP}} = 10^{15}$ m$^{-2}$, $T_{c}$ in FS Si is almost always greater than $T = 300$ K, with $T_{c}$ in FS Ge exceeding 300 K until roughly $E_{\perp} = 1.5$ V/\AA.
  However, FS Sn and encapsulated Si have by far the smallest $T_{c}$ at a given $E_{\perp}$, and we find that beyond about $E_{\perp} = 0.9$ V/\AA, $T_{c} < 300$ K.
  Analysis of the dependence of $T_{c}$ on $n_{\text{LP}}$ revealed that $T_{c}$ depends nearly linearly on $n_{\text{LP}}$, so that the critical concentration of LP to induce a BKT phase transition at $T_{c} = 300$ K does not exceed about $n_{\text{LP}} = 3.5 \times 10^{15}$ m$^{-2}$, even in encapsulated Si at high $E_{\perp}$.

  Finally, we examined in detail the optimization problem of trying to simultaneously maximize $\hbar \Omega_R$ and $T_{c}$, and presented the sets of conditions which could lead to a stable LP superfluid at $T = 300$ K, namely that the LP binding energy exceeds the thermal energy at temperature $T$, $E_{b,\text{LP}} > k_B T$, and that the BKT critical temperature $T_{c}$ exceeds the ambient temperature, $T_{c} > T$.
  Simultaneously satisfying these conditions yields a range of $E_{\perp}$ where a stable LP superfluid could form, with quantities such as $n_{\text{LP}}$, $N_X$, and $\gamma_{ex}$ acting as parameters.
  Since increasing $n_{\text{LP}}$ increases $T_{c}$, while increasing $E_{\perp}$ decreases $T_{c}$, it was found that increasing $n_{\text{LP}}$ would increase the maximum $E_{\perp}$ which could support an LP superfluid.
  Meanwhile, by increasing $E_{b,\text{LP}}$, the lower limit of $E_{\perp}$ is decreased, which can be achieved by either increasing $N_X$ or reducing $\gamma_{ex}$.

  It was determined that FS Si could support a stable LP superfluid at $T \geq 300$ K, intermediate $E_{\perp} \in [1.7,2.2]$ V/\AA, and relatively low LP concentrations, $n_{\text{LP}} = 10^{15}$ m$^{-2}$.
  The upper limit of $E_{\perp} = 2.2$ V/\AA~can be increased by increasing $n_{\text{LP}}$, but there is no convenient mechanism for decreasing the lower limit of $E_{\perp} = 1.7$ V/\AA, since $\gamma_{ex}$ is effectively fixed and increasing $N_X$ by stacking multiple FS Si monolayers on top of each other (in a sort of stacked bridge configuration) seems impractical.
  On the other hand, Type II Si is a more promising candidate for room-temperature superfluidity of LP, though it requires much higher LP concentration than FS Si.
  With $\gamma_{ex} = 5 \times 10^{13}$ s$^{-1}$, we find that the LP superfluid regime can be achieved for $n_{\text{LP}} \geq 2 \times 10^{15}$ m$^{-2}$ and $N_X \geq 4$ in the range $E_{\perp} \in [1.5~\text{V/\AA},1.8~\text{V/\AA}]$.
  As mentioned in Sec.~\ref{sec:polaritons}, based on experiments in TMDCs, we consider the possibility that $\gamma_{ex}$ in encapsulated Si could be much lower than in the FS Xenes.
  Taking $\gamma_{ex} = 10^{13}$ s$^{-1}$, we then predict that the superfluid phase can be achieved for $T > 300$ K in the range $E_{\perp} \in [1.2~\text{V/\AA},1.8~\text{V/\AA}]$, for $n_{\text{LP}} = 2 \times 10^{15}$ m$^{-2}$ and $N_X \geq 4$.
  By increasing the concentration to $n_{\text{LP}} = 3 \times 10^{15}$ m$^{-2}$, we predict the existence of the superfluid LP phase at $T > 300$ K for $N_X \geq 5$ in the range $E_{\perp} \in [1.0~\text{V/\AA},2.4~\text{V/\AA}]$.

  Let us now mention two important caveats to the three conditions for a stable LP superfluid, given in the previous section: the effect of temperature on LP concentration and lifetime, and the superfluid concentration at temperatures very slightly less than $T_{c}$.
  Since a significant fraction of particles in a gas at an average temperature $T$ would have kinetic energies that exceed $k_B T$, the LP binding energy should comfortably exceed 26 meV so that the population of LP is not significantly decreased due to thermal dissociation.
  To put it another way, in the low-temperature limit, it is often possible to experimentally correlate pump intensity and LP concentration \textendash{} the effect of increasing temperature would be to decrease the LP concentration for a given pump intensity.
  Furthermore, from Eq.~\eqref{eq:nsdensity}, the concentration of the superfluid component of the weakly interacting Bose gas of LP is given by the difference between the total concentration $n_{\text{LP}}$ and the critical concentration of the normal component, $n_{c}$.
  Therefore, the total LP concentration should exceed $n_{c}$ by an appreciable amount in order to obtain a non-negligible superfluid concentration.

  \section{\label{sec:conc}Conclusions}

  In this paper we demonstrated that the combination of tunable excitons (via a perpendicular electric field) in a tunable optical microcavity (via the cavity length) has the potential to give researchers unprecedented control over the Rabi splitting of polaritons as well as their collective properties such as the critical temperature of the BKT superfluid phase transition.
  Our results show that the properties of polaritons, especially the Rabi splitting, are highly sensitive to the cavity configuration (1 or 2 DBR) and the electric field.
  Indeed, we found that the significant increase in the effective cavity length in the 2 DBR configuration compared to the 1 DBR case strongly suppresses the Rabi splitting, to the point that the strong-coupling regime is only barely achieved in encapsulated Si at very high electric fields.
  However, for the 1 DBR configuration, the strong coupling regime can be achieved at small electric fields for the FS Xenes and moderate electric fields in encapsulated Si.
  We also considered stacking multiple Si monolayers on top of each other in the case of encapsulated Si \textendash{} for $N_X \geq 2$, the enhancement of the exciton-photon coupling constant $V$ is significant enough to drastically reduce the threshold $E_{\perp}$ for the onset of the strong coupling regime in encapsulated Si.
  Based on previous experiments on polaritons in TMDCs embedded in an optical microcavity, the excitonic inhomogeneous line broadening, $\gamma_{ex} = 5 \times 10^{13}$ s$^{-1}$, which we use as a baseline value in our calculations, also significantly reduces the Rabi splitting.
  It was shown in TMDCs that encapsulating the TMDC monolayer with $h$-BN very strongly suppresses the excitonic broadening \textendash{} by considering a similar reduction in $\gamma_{ex}$ in encapsulated Si compared to the FS Xenes, we find that the strong coupling regime in encapsulated Si could be achieved at quite low electric fields.

  Next, we closely analyzed the conditions for the formation of a superfluid of lower polaritons in the Xenes embedded in a microcavity.
  As was the case for the Rabi splitting, our results showed that $T_{c}$ is largest for the FS Xenes, but unlike the Rabi splitting, $T_{c}$ is inversely proportional to $E_{\perp}$.
  By analyzing the critical concentration for the onset of the BKT phase transition at fixed temperature, it was shown that an LP concentration of at most $n_{\text{LP}} = 3 \times 10^{15}$ m$^{-2}$ is required for even encapsulated Si to exhibit an LP superfluid at $T = 300$ K at very large $E_{\perp}$.
  The nearly linear relationship between $T_{c}$ and $n_{\text{LP}}$ for fixed $E_{\perp}$ is beneficial in terms of pushing for larger $n_{\text{LP}}$ in order to increase $T_{c}$.

  Finally, we considered in detail the feasibility of room-temperature superfluidity in FS Si and encapsulated Si in particular.
  It was established that a stable LP superfluid simultaneously requires (i) a large exciton binding energy, (ii) a large Rabi splitting (so that the LP do not dissociate due to thermal interactions), and (iii) a BKT critical temperature greater than or equal to room temperature at experimentally obtainable LP concentrations, for which we specify the condition $T_{c} \geq 300$ K.
  We note that because the exciton binding energy always exceeds the Rabi splitting at a given $E_{\perp}$, we focus our attention on the range of $E_{\perp}$ which satisfies conditions (ii) and (iii).
  It was found that a stable LP superfluid should be possible in FS Si for $E_{\perp} \in [1.7,2.2]$ V/\AA.
  Increasing $n_{\text{LP}}$ would increase the maximum $E_{\perp}$ in that range, but since $N_X = 1$ and $\gamma_{ex} = 5 \times 10^{13}$ s$^{-1}$ are effectively fixed for FS Si, there is no way to decrease the minimum $E_{\perp}$.

  On the other hand, in encapsulated Si we considered the conditions for room-temperature superfluidity of LP subject to the variation in three parameters: $n_{\text{LP}}$, $\gamma_{ex}$, and $N_X$.
  The additional adjustable parameters were shown to afford experimentalists much more freedom in constructing a device which could exhibit room-temperature superfluidity of LP in a wide range of $E_{\perp}$ based on the choice of $N_X$ and the value of $\gamma_{ex}$ in encapsulated Si.
  If, in encapsulated Si, the excitonic line broadening $\gamma_{ex} = 5 \times 10^{13}$ s$^{-1}$, we predict the existence of a stable LP superfluid at $T = 300$ K between $E_{\perp} \in [1.5, 1.8]$ V/\AA, for $N_X \geq 3$ and $n_{\text{LP}} \geq 2 \times 10^{15}$ m$^{-2}$.
  For $\gamma_{ex} = 10^{13}$ s$^{-1}$, we find that the range of $E_{\perp}$ increases to $E_{\perp} \in [1.2,1.8]$ for $N_X = 3$ and $n_{\text{LP}} = 2 \times 10^{15}$ m$^{-2}$.

  Our results indicate that the Xenes warrant intensive study alongside more well-studied 2D materials such as graphene and the TMDCs.
  Their tunable nature via and external electric field allows for highly flexible manipulations of excitons, which can be naturally extended to polaritons in an open microcavity.
  The prospect of room-temperature superfluidity of LP in encapsulated Si is especially noteworthy because of how closely such a set-up would resemble current experimental work in the TMDCs.
  Therefore, much more work remains to be done in studying polaritons in the Xenes in optical microcavities.

\appendix
\section{\label{app:hopcoeff}Hopfield Coefficients}

  The exciton-photon Hamiltonian of Eq.~\eqref{eq:cavityham} can be diagonalized by means of a linear unitary transformation~\cite{Deng2010},

  \begin{align}
	  \hat{p}_{\mathbf{k}} & = X_{\mathbf{k}} \hat{b}_{\mathbf{k}} + C_{\mathbf{k}} \hat{a}_{\mathbf{k}} \nonumber\\
	  \hat{u}_{\mathbf{k}} & = -C_{\mathbf{k}} \hat{b}_{\mathbf{k}} + X_{\mathbf{k}} \hat{a}_{\mathbf{k}},
	  \label{eq:polbasislintrans}
  \end{align}
  where $\hat{p}_{\mathbf{k}}$ $(\hat{p}_{\mathbf{k}}^{\dag})$ and $\hat{u}_{\mathbf{k}}$ $(\hat{u}_{\mathbf{k}}^{\dag})$ are the Bose annihilation (creation) operators of lower and upper polaritons, respectively, and the coefficients $X_{\mathbf{k}}$ and $C_{\mathbf{k}}$ are known as the Hopfield coefficients~\cite{Hopfield1958}, and are given by~\cite{Deng2010}:

  \begin{align}
	  \lvert X_{\mathbf{k}} \rvert^2 = \frac{1}{2} \left( 1 + \frac{\Delta E(\mathbf{k})}{\sqrt{\left( \Delta E(\mathbf{k}) \right)^2 + \left( 4 \hbar V \right)^2}} \right) \nonumber \\
	  \lvert C_{\mathbf{k}} \rvert^2 = \frac{1}{2} \left( 1 - \frac{\Delta E(\mathbf{k})}{\sqrt{\left( \Delta E(\mathbf{k}) \right)^2 + \left( 4 \hbar V \right)^2}} \right),
	\label{eq:hopfielddefs}
  \end{align}
  where $\Delta E(\mathbf{k}) = E_{ph} (\mathbf{k}) - E_{ex} (\mathbf{k})$.
  Let us further note that $\lvert X_{\mathbf{k}} \rvert^2$ and $\vert C_{\mathbf{k}} \rvert^2$ give the excitonic and photonic fractions of lower polaritons, respectively, and $\lvert X_{\mathbf{k}} \rvert^2 + \lvert C_{\mathbf{k}} \rvert^2 = 1$.

\section{\label{app:v}}

  In Ref.~\onlinecite{Savona1995}, the authors considered polaritons in semiconductor quantum wells in an optical microcavity, and obtained the following expression for $V$ in an optical microcavity with one active layer,

  \begin{equation}
	  V = \left[ \frac{1 + \sqrt{R}}{\sqrt{R}} \frac{c \Gamma_0}{\sqrt{\epsilon_{cav}} L_{eff}} \right]^{1/2}.
	  \label{eq:SavV}
  \end{equation}
  In Eq.~\eqref{eq:SavV}, $\Gamma_0$ is the decay rate of the exciton amplitude, given by~\cite{Dufferwiel2015}:

  \begin{equation}
	\Gamma_0 = \left( \frac{\pi k e^2}{m_0 c \sqrt{\kappa}} \right) \left( \frac{2 \lvert P_{cv} \rvert^2}{m_0 E_{ex}} \right) \lvert \psi(\rho=0) \vert^2,
	\label{eq:duffergamma}
  \end{equation}
  where $m_0$ is the electron rest mass, $P_{cv}$ is the dipole transition matrix element given in terms of the momentum operator of the exciton-forming optical transition associated with the transition energy $E_{ex}$, and $\psi(\rho=0)$ is the value of the direct exciton relative-motion eigenfunction at $\rho=0$.

  In Ref.~\onlinecite{Vasilevskiy2015}, the authors approximated the momentum dipole matrix element as $\lvert P_{cv} \rvert^2 \approx 2 m_0^2 v_F^2$.
  We also note that it is well known that $V \propto \sqrt{N}$, where $N$ is the number of quantum wells~\cite{Deveaudbook} or TMDC monolayers~\cite{Dufferwiel2015}, therefore we add a factor $\sqrt{N_X}$ to our expression for $V$, where $N_X$ is the number of stacked Xene monolayers.

  Finally, combining Eqs.~\eqref{eq:SavV},~\eqref{eq:duffergamma}, the approximate expression for $\lvert P_{cv} \rvert^2$, and the proportionality $V \propto \sqrt{N_X}$, we obtain Eq.~\eqref{eq:SavVplug}.

  \section{Acknowledgements}

  This work is supported by the U.S. Department of Defense under Grant No. W911NF1810433.

\bibliography{sgspublish}

\end{document}